\documentclass[11pt]{article}

\usepackage{enumerate}

\usepackage{mathtools}

\usepackage{needspace}

\AtBeginDocument{
	\addtocontents{toc}{\normalsize}%
}
\ifdefined\XeTeXversion\else\pdfoutput=1\fi

\usepackage{amsmath,amssymb,amsfonts,amsthm,amscd,mathrsfs}

\usepackage[most]{tcolorbox}

\theoremstyle{plain}
\newtheorem{theorem}{Theorem}[section]
\newtheorem{lemma}[theorem]{Lemma}
\newtheorem{proposition}[theorem]{Proposition}

\theoremstyle{definition}

\newtheorem{problem}[theorem]{Problem}

\theoremstyle{remark}

\tcolorboxenvironment{problem}{
  boxrule=0.4pt, colframe=black!55, colback=black!2,
  sharp corners, left=6pt, right=6pt, top=4pt, bottom=4pt,
  before skip=12pt, after skip=14pt
}
\usepackage{esvect}
\usepackage{xcolor}
\definecolor{darkblue}{rgb}{0.1,0.1,.7}
\usepackage[]{version}
\usepackage[]{graphicx}
\usepackage{wrapfig}
\usepackage[]{latexsym}
\usepackage{geometry}
\usepackage[all,cmtip]{xy}

\usepackage[margin=10pt,font=small,labelfont=bf]{caption}
\usepackage{ifthen}
\usepackage{soul}
\usepackage{tikz}
\usepackage{array,mathrsfs,amsfonts,yfonts,dsfont,bbm,colonequals}

\usepackage{dsfont}
\usepackage{cite}
\usepackage{xspace}
\usepackage{empheq}
\usepackage{extarrows}

\usepackage{xcolor}
\usepackage[colorlinks, linkcolor=darkblue, citecolor=darkblue, urlcolor=darkblue, linktocpage]{hyperref} 
\hypersetup{
  pdftitle={Existence of extremal solutions in the 1d conformal bootstrap},
  pdfauthor={Filip Her\v{c}ek and Nat Levine}
}

\usepackage{subcaption}
\usepackage[utf8]{inputenc}
\usepackage{float, graphicx}

\normalsize

\numberwithin{equation}{section}

\newcommand{\be}{\begin{equation}}
\newcommand{\ee}{\end{equation}}
\newcommand{\bea}{\begin{eqnarray}}
\newcommand{\eea}{\end{eqnarray}}
\newcommand{\ba}{\begin{equation}\begin{aligned}}
\newcommand{\ea}{\end{aligned}\end{equation}}

\newcommand{\ud}{\mathrm d}

\newcommand{\ket}[1]{|#1\rangle}

\def\g{\gamma}

\def\a{\alpha}
\def\e{\epsilon}
\def\k{\kappa}
\def\b{\beta}

\def\D{\Delta}

\def\l{\lambda}

\def\la{\langle}

\def\Re{{\rm Re}}

\usepackage{ulem}

\def\la{\label}
\def\rf{\eqref}
\usepackage{cancel}
\def\p{\phi}

\usepackage{blindtext}

\makeatletter
\newcommand{\global@insert}[2]%
{\bgroup
  \setbox\@tempboxa=\box#1
  \global\setbox#1=\vbox{\unvbox\@tempboxa #2}
\egroup}

\usepackage{changepage}

\newcommand{\ha}{\tfrac{1}{2}}

\newcommand{\no}{\nonumber}

\begin{document}
\normalem
	
	\vspace*{-.4in} \thispagestyle{empty}
	\vspace{3cm} {\Large
		\begin{center}
			{\bf Existence of extremal solutions in the 1d conformal bootstrap
}\\
	\end{center}}
	\vspace{1.5cm}
	\begin{center}
		{\bf Filip Her\v{c}ek \quad and \quad  Nat Levine}
		{\vspace{1.3cm}
			
			{\em Institute  for  Theoretical  Physics,  University  of  Amsterdam,\\ \vspace{0.1cm}
            PO  Box  94485,  1090  GL, Amsterdam, The Netherlands}

			\normalsize
		}
		
	\end{center}
	\vspace{0.5cm}
	\begin{center}
		{\texttt{f.hercek@uva.nl}, \quad \texttt{n.j.levine@uva.nl} 
		}
		\\
	\end{center}
	
	\vspace{13mm}
	
	\begin{abstract}
 \normalsize
 \vspace{2mm}
 Extremal solutions and functionals play an important role in the conformal bootstrap. These are the solutions that saturate unitarity bounds, and the dual functionals that encode the optimal bounds. While they have fundamental theoretical implications, their existence has not been rigorously established. This is particularly subtle in the setting of the full infinite-dimensional bootstrap, i.e.\ the limit of zero numerical truncation. In this article, we prove that extremal solutions exist for OPE maximisation problems in 1d CFT. These are a large class of generically interacting, unitary, exact solutions of the crossing equation that maximise an OPE coefficient. We also prove that these problems have `zero duality gap', meaning there exist functionals arbitrarily close to extremality. Our approach is to formulate the bootstrap as a convex optimisation problem in a Hilbert space and apply the Fenchel-Rockafellar theorem. 
	\end{abstract}
	\vspace{2in}

	\newpage
	
	\tableofcontents

\section{Introduction}

The idea of the conformal bootstrap \cite{Ferrara:1973yt,Polyakov:1974gs,Rattazzi:2008pe} is to impose basic consistency conditions on correlation functions in conformal field theory (CFT) and extract the consequences for the dynamical data: the scaling dimensions and OPE coefficients. In its modern form \cite{Rattazzi:2008pe,Poland:2018epd,Rychkov:2023wsd}, this consists of imposing \textit{crossing} (equivalence of OPE expansions in different channels) and \textit{unitarity} (reality of OPE coefficients). 

This programme has been remarkably successful. One message is that conformal invariance is very constraining, and there are `not many' solutions, particularly in $d>2$. A pioneering series of works on the elusive 3d Ising model have led to rigorous bounds that constrain its low-lying CFT data to tiny `islands', determining them with many digits of precision~\cite{ElShowk:2012,ElShowk:2014,Kos:2014bka,Kos:2016,Chang:2024whx}. For a given finite system of correlators, the axioms of crossing and unitarity can be written as a \textit{semi-definite} system, i.e.\ a set of affine and positive-semi-definite constraints on matrices~\cite{Kos:2014bka,SimmonsDuffin:2015}. Then the space of allowed CFT data can be explored numerically by semi-definite programming (SDP), that is, maximising linear functions of these matrices. Increasingly advanced techniques are available to implement this method and search the space of scaling dimensions~\cite{SimmonsDuffin:2015,Reehorst:2021ykw,Liu:2023elz}.

In fact, these methods usually use a `dual' SDP, rather than the `primal' one described above. This allows one to rigorously exclude regions of parameter space, rather than constructing candidate solutions. The dual formulation is in terms of linear functionals, which can be applied to the crossing equation to digest its kinematic dependence. Functionals with certain positivity properties imply bounds on unitary CFT data, and one can then numerically optimise these bounds.

\begin{figure}[!t]
    \centering
    \begin{minipage}[t]{0.48\textwidth}
        \centering
        \includegraphics[width=\linewidth]{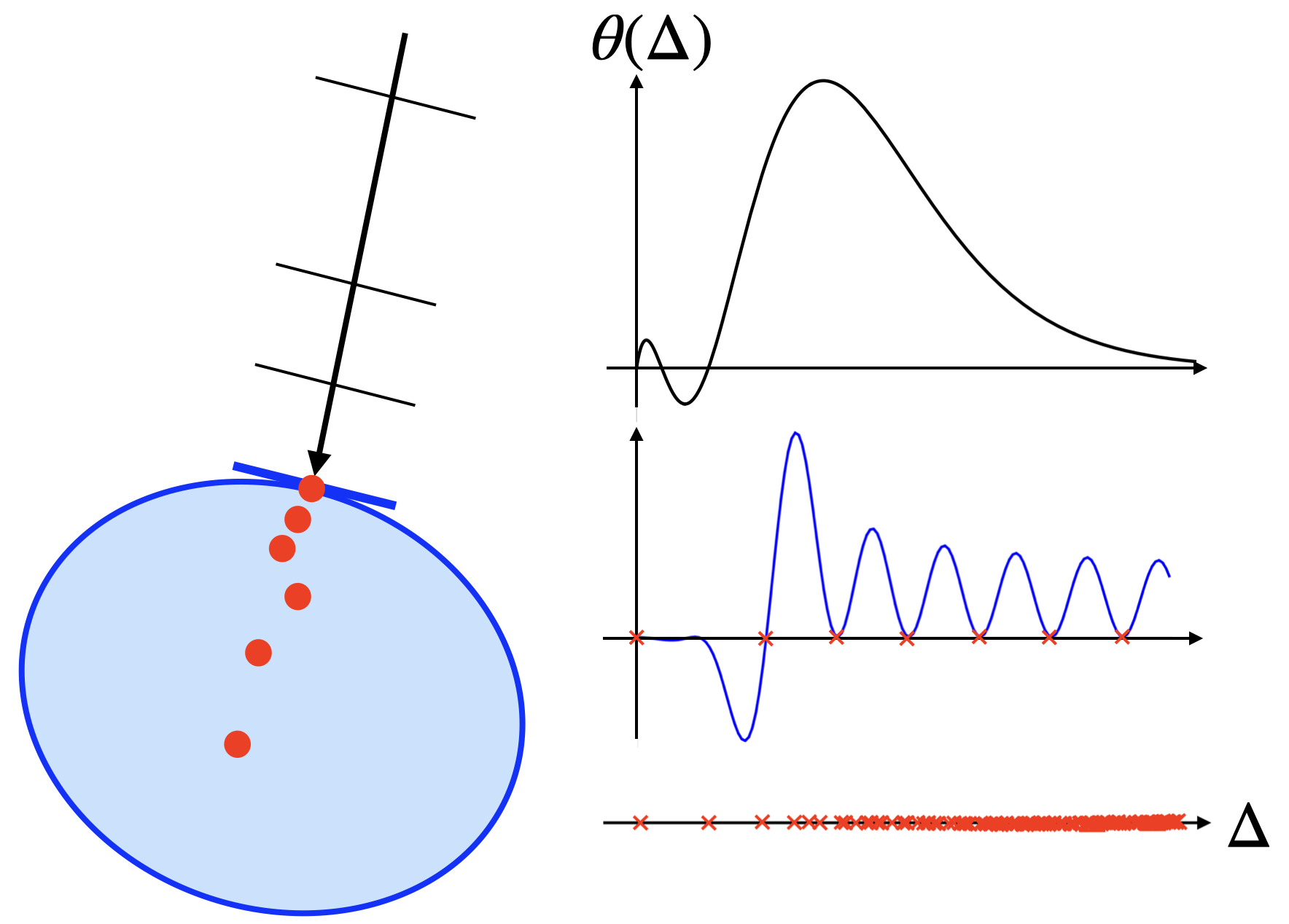}
        \captionsetup{margin=0pt}
        \caption{\small The expected lore for extremal solutions and functionals. Generic positive functionals (black) have few zeros. Generic solutions (red) in the interior of the allowed region have dense, chaotic spectra. At the boundary, extremal functionals have maximal sets of zeros and extremal solutions have minimally dense spectra, and the two coincide.}
        \label{fig:lore}
    \end{minipage}\hfill
    \begin{minipage}[t]{0.38\textwidth}
        \centering
        \includegraphics[width=\linewidth]{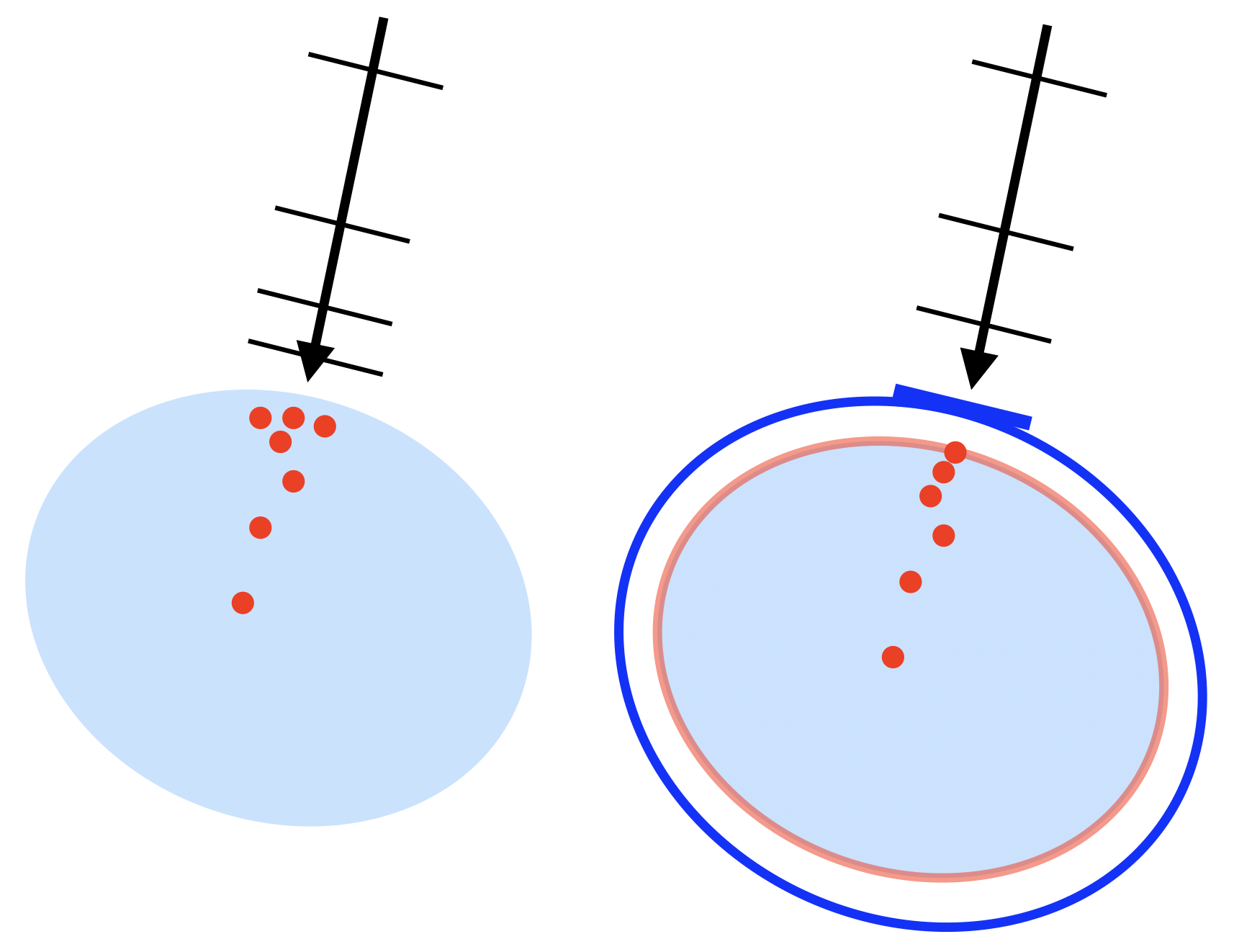}
        \caption{\small A sketch of possible failure modes for existence and duality of extremal functionals and solutions. Left: non-attainment of optimal values. Right: attainment but non-zero duality gap.}
        \label{fig:failure_modes}
    \end{minipage}
\end{figure}

A robust finding in numerical studies is the appearance of \textit{extremal solutions} and \textit{extremal functionals} with special properties. Extremal solutions are correlators that \textit{saturate} the constraints, i.e.\ they maximise some quantity subject to unitarity and crossing. Meanwhile, extremal functionals are those that give an optimal bound. The idea advanced in the Extremal Functional Method of \cite{ElShowk:2013} and elaborated in \cite{ElShowk:2014,El-Showk:2016mxr} is that, at the boundary of the allowed region, extremal solutions and functionals appear and are dual to each other. This means the spectrum of the extremal solution is supported on the set of zeros of the extremal functional. This makes them very special: while generic solutions would have a much denser spectrum, extremal ones are expected to have the minimal density of operators; and while generic functionals would have far fewer zeros, extremal ones are expected to have the maximal number: see Figure~\ref{fig:lore}. Extremal solutions have played an important role in various recent ideas in the bootstrap \cite{Afkhami-Jeddi:2019zci,Erramilli:2026esf}, and have been used as proxies to study features such as Regge trajectories in nearby physical CFTs \cite{Simmonsduffin:2017}.

As we shall review, SDPs are instances of \textit{convex optimisation} problems, for which the existence of dual pairs of optimisers can be guaranteed under certain conditions. Similar arguments using convexity were made in \cite{ElShowk:2013}, which apply straightforwardly in the finite-dimensional setting, i.e.\ at finite numerical truncation $\Lambda<\infty$, where $\Lambda$ is the number of functionals in the truncation. This work is inspired by the question of whether extremal solutions and functionals persist and converge in the limit $\Lambda\to\infty$ when numerics have fully converged:

\begin{adjustwidth}{3em}{3em}
\vspace{0.2cm}
\textit{Do extremal solutions and functionals necessarily exist in full (infinite-dimensional) conformal bootstrap problems? What are their properties? Are they unique?}
\vspace{0.2cm}
\end{adjustwidth}

\noindent  These properties are far from obvious. There are several failure modes: for example, it could be that solutions and functionals come arbitrarily close to extremality but never actually attain it. Even if they do attain extremality, it could be that their extremal values do not coincide: this is known as a non-zero \textit{duality gap}. See a sketch of these possibilities in Figure~\ref{fig:failure_modes}.

These are naturally subtle questions, with all of the complications of infinite dimensions. However, there is quite a lot of positive evidence. Beginning in 1d CFT, Maz\'a\v{c}~\cite{Mazac:2016qev} and Maz\'a\v{c} and Paulos~\cite{Mazac:2018mdx,Mazac:2018ycv} wrote down complete bases of analytic functionals.\footnote{Analytic functionals have later been generalised to higher dimensions~\cite{Paulos:2019gtx,Mazac:2019shk,Caron-Huot:2020adz}.} These are `analytic' in the sense that they encode optimal bounds in the infinite-dimensional problem. Each one is the extremal functional for a bootstrap problem saturated by a Generalized Free Boson (GFB) or Generalized Free Fermion (GFF) correlator---so we may call them `free' functionals. Remarkably, in this case, numerical functionals are seen to converge  in a well-behaved manner to the analytic functionals as $\Lambda\to\infty$. It was not obvious a priori that this had to happen, and that the limit would be a healthy functional that satisfies the swapping property \cite{Qiao:2017lkv} (i.e.\ commutes with the OPE expansion).

No such construction of analytic functionals dual to \textit{interacting} extremal solutions has been achieved.\footnote{Let us mention a related bootstrap problem for local bulk operators in AdS \cite{LP}, which does not have any positivity condition. In that case one can import the notion of extremality defined purely in terms of sparseness, but without any bounds. In that setting it was possible to explicitly write down analytic functionals and solutions with arbitrary sparse spectra \cite{part2}.} However, a family of interacting extremal functionals and dual extremal solutions were found numerically \cite{Paulos:2019fkw} for an OPE-maximisation problem in 1d that interpolates between GFB and GFF solutions (the `$\phi^4$ flow', also discussed in several other works \cite{Mazac:2018ycv,part2,Ghosh:2025sic}). Our main result rigorously proves the existence of these optimal solutions, as well as those for a broad class of generalised OPE-max problems.

\paragraph{Main result.} Consider a 4-point correlator of identical scalar primaries $\langle \phi \phi \phi \phi \rangle$ of dimension $\Delta_\phi>0$ in 1d CFT. Its conformal-invariant part is a function of one cross-ratio $G(z)$. Fix any `gap' $\Delta_*\geq \sqrt{2}\Delta_\phi$ and any low-lying `source' $S(z)$, that is the choice of blocks and OPE coefficients appearing below the gap:
\be
G(z) = S(z) + \sum_{\Delta\geq \Delta_*} a_\Delta \, G_\Delta(z)  \ , \qquad S(z) = \sum_{0\leq \Delta< \Delta_*} a_\Delta^{\rm fixed} \, G_\Delta(z) \ , \label{eq:gapSource}
\ee
where $G_\Delta(z)$ are conformal blocks. As we shall explain below, the (squared) OPE coefficients $a_\Delta = \lambda_{\phi \phi \D}^2$ are taken here to be a Radon measure over $\Delta$, and may in general contain discrete and/or continuous contributions. In this setup, our result can be stated as follows:

\vspace{0.2cm}
\begin{theorem}[OPE-max primal attainment]\label{thm:main}
   For a given gap and source \eqref{eq:gapSource}, consider the problem of maximising the OPE coefficient $a_{\Delta_*}$ of the operator of dimension $\Delta_*$,
       \begin{align}
  &{\rm Maximise} && \hspace{-3.5cm} a_{\Delta_*} \no\\
  &{\rm subject\ to} && \hspace{-3.5cm} {\rm crossing\  and\ unitarity} \ . \no
\end{align}
   If there exists any feasible solution, then there is an optimal one.
\end{theorem}
\vspace{0.1cm}
The hypothesis of feasibility depends on the choice of gap and source. In the simplest case of the `$\phi^4$ flow' mentioned above, the source is just the identity operator, $S(z)=G_0(z)$.  Parametrised by the choice of gap $\Delta_*$, these solutions are believed to smoothly interpolate between GFB at $\Delta_*=2\Delta_\phi$ (known to saturate OPE maximisation) and GFF at $\Delta_*=2\Delta_\phi+1$ (the maximal value of the gap $\D_*$). We know that this problem is feasible for all values $\Delta_*\leq 2\Delta_\phi+1$: this is automatic because of existence of the GFF solution, whose lowest exchanged operator is at $\Delta=2\Delta_\phi+1$. Hence the theorem proves the existence of all of these optimal solutions unconditionally.

To prove this theorem, we shall model the space of correlation functions $G(z)$ as a \textit{Bergman space} $A^2$, that is a space of holomorphic functions on a domain with a finite (weighted) $L^2$ norm.\footnote{It is interesting to note that Bergman spaces have previously been used in the CFT context as models of individual representations of the conformal group \cite{Ammon:2024axd,perelomov1972coherent}. We are not aware of any direct link between these uses.} We formulate the bootstrap problem as a convex optimisation problem in this infinite-dimensional space, and use the Fenchel-Rockafellar duality theorem to prove that it attains its maximum if it is feasible. We also show that there is \textit{zero duality gap} between this primal problem and a dual one for functionals defined in the same space, meaning their optimal values are equal. We prove that this Bergman space optimisation problem is completely equivalent to the physical problem stated above. The extremal solutions are not formal objects: they are positive combinations of conformal blocks that solve the crossing equation. 

We emphasise that this study addresses the full infinite-dimensional bootstrap problem with $\Lambda=\infty$. This is the limit when numerics have fully converged. Let us briefly comment on what is already known about finite truncations $\Lambda<\infty$ and optimal attainment in that context. At finite $\Lambda$, the functionals imposing crossing are restricted to a finite-dimensional subspace, while the primal problem of solving those crossing constraints is still infinite-dimensional. Focusing still on a single 4-point correlator, one can show that the dual optimum is attained at finite $\Lambda$, assuming primal and dual feasibility; a similar statement was made in \cite{ElShowk:2013}. In contrast, the primal optimum is \textbf{not} always obtained: one can construct simple counterexamples where operators run off to infinity as one approaches optimality (see related discussion in \cite{El-Showk:2016mxr,Erramilli:2026esf,Huang:2019xzm}). For multiple correlators, which are genuine SDPs rather than linear programs, even dual attainment becomes dependent on a non-trivial \textit{Slater condition}---discussed in the related context of SDPB implementation in~\cite{SimmonsDuffin:2015,Liu:2023elz,Reehorst:2026xnt}.

This paper focuses on 1d CFT for kinematical simplicity. This can be seen as a toy model for higher-d CFT, which we expect to be amenable to the same techniques. There are also many interesting and important 1d CFTs in their own right, including conformal line defects (see \cite{Gaiotto:2013nva,Liendo:2018ukf,Giombi:2017cqn,Cavaglia:2021bnz}), long-range CFTs (see \cite{Fisher:1972zz,Paulos:2015jfa,Benedetti:2025nzp,Ghosh:2026gku}) and  the boundary correlators of QFTs in AdS$_2$ (see \cite{QFTinAdS,Mazac:2016qev,Antunes:2021abs,Antunes:2024hrt,Antunes:2025iaw}). While these theories are non-local in the sense of not having a stress tensor, they are still perfectly healthy, satisfy conformal invariance, and solve the conformal bootstrap. It is also worth noting that any CFT in general dimension has a 1d CFT subsector with all operators lying on a circle, which is constrained to satisfy all of the 1d CFT bootstrap bounds.

\paragraph{Significance of extremal solutions.}
Let us mention some other matters that motivate us to study extremal solutions.

First, extremal solutions suggest the appealing possibility of a \textbf{hybrid bootstrap} method. Consider the islands produced through the numerical conformal bootstrap. These are the high-value outputs of very intensive SDP numerics, for example of order $10^7$ CPU hours for the 3d Ising bounds in~\cite{Chang:2024whx}. Suppose the above questions could be very well understood, and one knew that the boundary of the island was covered by extremal solutions, which were the unique solutions with a particular type of spectra. Then one may imagine dispensing with the SDP numerics, and instead performing a direct numerical search for extremal solutions. This would surely be much  faster and cheaper---and would open the door to more difficult settings, for example in 4d CFT, where numerical convergence is an even more pressing bottleneck. A programme of this type has already been explored by Ghosh, Paulos and Suchel in 1d CFT \cite{Ghosh:2025sic}. Those authors numerically solve for extremal solutions and functionals in a manner that is demonstrably efficient and versatile. One of the goals of this work is to further justify this hybrid bootstrap approach by rigorously establishing the existence and properties of extremal solutions---which were previously taken as assumptions.

Second, a longstanding question is whether the islands produced by the conformal bootstrap can shrink to \textbf{zero size} for a finite system of correlators. This would mean the bootstrap could truly isolate physical CFTs of interest such as 3d Ising. This hope was encouraged by the fact that the kinks from the single correlator bootstrap \cite{ElShowk:2012,ElShowk:2014} were replaced by very small islands when upgrading to multiple correlators \cite{Kos:2014bka}. A plausible belief was that those islands would shrink to zero size either in the limit of increasing numerical truncation, or for some larger but finite collection of correlators. The present expectations about extremal solutions weigh against that hope. If, as expected, extremal solutions for finite systems always have sparse spectra,\footnote{See the study of extremal solutions in systems of multiple correlators in \cite{Ghosh:2023wjn}.} and if 3d Ising has a chaotic, dense spectrum as expected, this would rule it out from being extremal for a finite system of correlators. Rather, one would expect that the islands may shrink to zero size only in the limit of infinitely many correlators. In order to clarify these questions, it is naturally of great importance to understand the properties of extremal solutions.

\paragraph{Structure of the paper.} We begin in \textbf{Section~\ref{sec:setup}} by setting up the 1d conformal bootstrap, and informally stating the problem of OPE maximisation, as well as its formulation in terms of dual functionals. In \textbf{Section~\ref{sec:Convex-Hilbert}}, we turn to a formal treatment of the problem and prove our main result: existence of extremal solutions for feasible OPE maximisation problems in 1d CFT. 

We discuss these results and future directions in \textbf{Section~\ref{sec:disc}}. Several properties of conformal blocks are derived in \textbf{Appendix~\ref{App:Blocks}}.

\section{1d conformal bootstrap and OPE maximisation}\label{sec:setup}
In this section we shall review 1d CFT and the conformal bootstrap, in particular the OPE maximisation problem.

\subsection{1d conformal bootstrap}
This work focuses on 4-point correlation functions of identical scalar primaries $\langle \phi \phi \phi \phi\rangle$ in 1d CFT. In 1 dimension, the conformal-invariant part of this correlator is a function of a single conformal cross-ratio $z$,
\be
\left\langle \phi(x_1) \phi(x_2) \phi(x_3) \phi(x_4) \right\rangle = \frac{1}{|x_{13}x_{24}|^{2\D_\p}} \, G(z)  \ , \qquad z\equiv \frac{x_{12}x_{34}}{x_{13}x_{24}} \ , 
\ee
where $\Delta_\phi>0$ is the conformal weight of $\phi$.

The correlator admits a convergent OPE expansion around $z=0$, under which $G(z)$ is expanded into conformal blocks $G_\Delta(z)$, 
\be
G(z) = \sum_\Delta a_\Delta \, G_\Delta(z) \ , \qquad G_\Delta(z) = z^{\Delta-2\Delta_\phi}\, _2F_1(\D,\D;2\D;z) \ , \la{OPE}
\ee
where, in particular, the identity block is $G_{0} = z^{-2\D_{\p}}$. The sum ranges over the unitary representations $\Delta\geq 0$ that may be exchanged in the OPE. The dynamical variables are the OPE coefficients $a_\Delta=\l_{\phi\phi\Delta}^2$ as well as the values of $\Delta$ that appear in the sum. The identity operator appears with coefficient $a_0=1$ by normalization convention of $\phi$. In physical examples, this sum is almost always discrete, but in our analysis we will also allow for possible continuous contributions.

Unitarity imposes that the OPE coefficients  $\l_{\phi\phi\Delta}$ are real, and hence their squares $a_\Delta=\l_{\phi\phi\Delta}^2$ non-negative,
\be
\text{Unitarity:}\quad a_\Delta\geq 0 \ . \label{unit}
\ee
Since the choice of OPE expansion was arbitrary, a relabelling of the points leads to the invariance of this function under $z\mapsto 1-z$, called crossing symmetry,
\be
\text{Crossing:}\quad G(z) = G(1-z) \ . \label{cross}
\ee
One can decompose the crossing equation using the OPE in terms of \textit{crossing vectors} $F_\Delta(z):= G_\Delta(z)-G_\Delta(1-z)$, the differences between direct- and crossed-channel conformal blocks,
\be
\sum_\Delta a_\Delta \, F_\Delta(z) = 0 \ . \label{crossblocks}
\ee
In the conformal bootstrap, one imposes unitarity and crossing as axioms and studies the space of solutions.

\begin{figure}[t]
    \centering
\includegraphics[width=\linewidth]{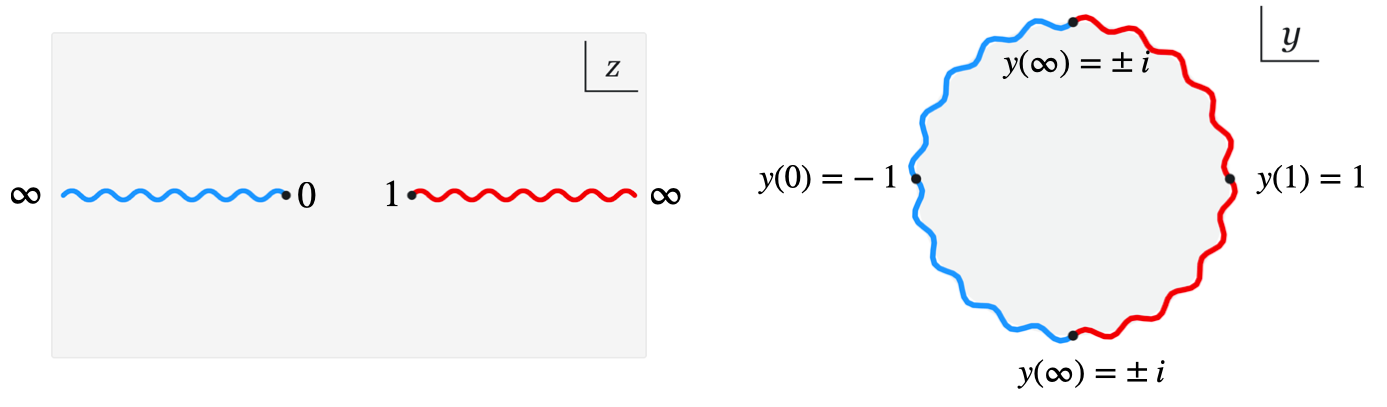}
    \caption{The domains of the conformal cross ratio $z$ and the Zhukovsky variable $y$, related to $z$ by \rf{Zhuk}.}
    \label{fig:yvar}
\end{figure}

Physically, the co-ordinate $z$ is valued in $(0,1)$. However, it is customary to complexify it, and the OPE \rf{OPE} converges absolutely in the region $z\in \mathbb{C}\setminus\left[1,\infty\right)$, with a branch cut at $(-\infty,0]$ \cite{Pappadopulo:2012jk}. As a result, the correlator $G(z)$ is analytic and the crossing equation \rf{cross} is valid on the entire cut plane $\mathbb{C}\setminus((-\infty,0]\cup[1,\infty))$. This was originally argued by changing to a `radial' variable $\rho$ on the cut unit disc \cite{Hogervorst:2013sma}, 
\be\begin{aligned}
&\rho(z): \ \mathbb{C}\setminus\left((-\infty,0]\cup[1,\infty)\right) \to \mathbb{D}\setminus(-1,0] \ ,  \qquad \qquad  z(\rho ) :=\frac{4 \rho}{(1+\rho)^2} \ ,
\end{aligned}\ee
where $\mathbb{D}$ denotes the open unit disc in $\mathbb{C}$. While this map fixes $0$ and sends the other OPE channels $z=1,\infty$ to the boundary of the disc, we shall favour the Zhukovsky variable $y$ of \cite{Mazac:2016qev} 
\be\begin{aligned}
&y(z): \ \mathbb{C}\setminus\left((-\infty,0]\cup[1,\infty)\right) \to \mathbb{D} \ ,  \qquad \qquad  z(y ) :=\frac{(1+y)^2}{2(1+y^2)} \ , \la{Zhuk}
\end{aligned}\ee
which  treats all channels equally and maps $z=0,1,\infty$ to the boundary of the disc (see Figure~\ref{fig:yvar}),
\be
y(0)=-1 \ , \qquad y(1)=1 \ , \qquad y(\infty) = \pm i  \ . \label{eq:crossrel}
\ee

Let us record some properties of conformal blocks that will be useful later. It is well known that they satisfy `radial dominance' in the radial co-ordinate, $|G_\Delta(\rho)|\leq |\frac{z(|\rho|)}{z(\rho)}|^{2\Delta_\phi}G_\Delta(|\rho|)$.\footnote{We abuse notation by writing $G_\Delta(\rho) \equiv G_\Delta(z(\rho))$, etc.} This is because (after stripping off $z^{-2\Delta_\phi}$) their expansion around $\rho=0$ has positive coefficients \cite{Hogervorst:2013sma}. A little work reveals the same property to be satisfied in the Zhukovsky variable. 
\begin{lemma}[Radial dominance in $y$]\label{lemma:RadDom}For $\Delta_\phi>0$ and $\Delta\geq0$, the conformal block $G_\Delta$ satisfies
\be
|G_\Delta(y)|\leq \left|\frac{z(|y|)}{z( y)}\right|^{2\Delta_\phi}G_\Delta(| y|) \  , \qquad y \in \mathbb{D} \ .
\ee
\end{lemma}
\noindent This lemma is proven in Appendix~\ref{App:Blocks}, together with the following three lemmas on the positivity properties on conformal blocks.
\begin{lemma}\label{lemma:ratBlocks}
    For $\Delta_\phi>0$ and $\Delta>0$, conformal blocks are strictly positive on $0<z<1$ and satisfy 
    \be
\frac{G_\Delta(z_1)}{G_\Delta(z_2)} \, \overset{\Delta\to\infty}\longrightarrow\,  0 \qquad {\rm if } \ \ \ \  0<z_1<z_2<1 
    \ee
    with the decay being exponential.
\end{lemma}
\noindent Even after passing to crossing vectors, some positivity properties are still preserved:
\begin{lemma}\label{lemma:crossPos}
        For $\Delta\geq 2\Delta_\phi>0$, crossing vectors are strictly positive on $\frac 12 < z<1$ and satisfy
    \be
(1-\kappa_{z}) G_\Delta(z) \leq F_\D(z) \leq G_\Delta(z)  \ , \quad \qquad \frac 12 < z<1 \ ,
    \ee
    where $\kappa_z \in (0,1)$ is independent of $\Delta$.
\end{lemma}
The positivity of $F_\Delta(z)$ on $\ha<z<1$ fails for $\D<2\Delta_\phi$. However, as long as $\Delta\geq \sqrt{2}\D_\phi$, one still preserves positivity on some window $(\tfrac12,\tfrac12+\e]$, or equivalently positivity of the derivative at $z=\tfrac12$:
\begin{lemma}\label{lemma:derivativePos}
    For $\Delta\geq \sqrt{2}\Delta_\phi>0$, crossing vectors satisfy
    \begin{equation}
        F_\Delta'(z=\tfrac12)>0 \ .
    \end{equation}
\end{lemma}

\begin{figure}[!t]
    \centering
    \settoheight{\dimen0}{\includegraphics[width=0.60\textwidth]{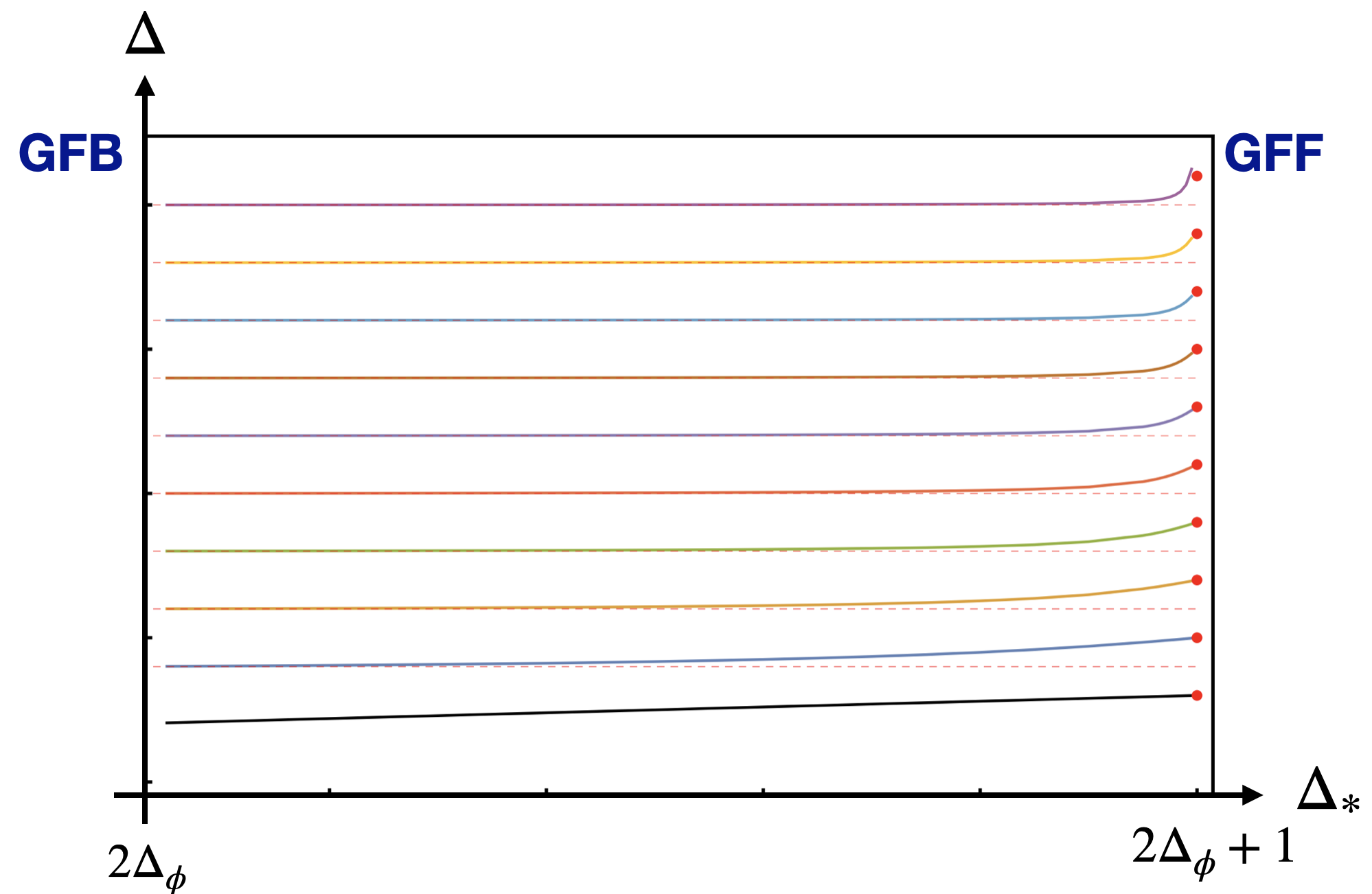}}
    \edef\phiFourPanelHeight{\the\dimen0}
    \begin{subfigure}[t]{0.4\textwidth}
        \centering
        \raisebox{\dimexpr(\phiFourPanelHeight-\height)/2\relax}[\phiFourPanelHeight][0pt]{%
            \includegraphics[width=\linewidth]{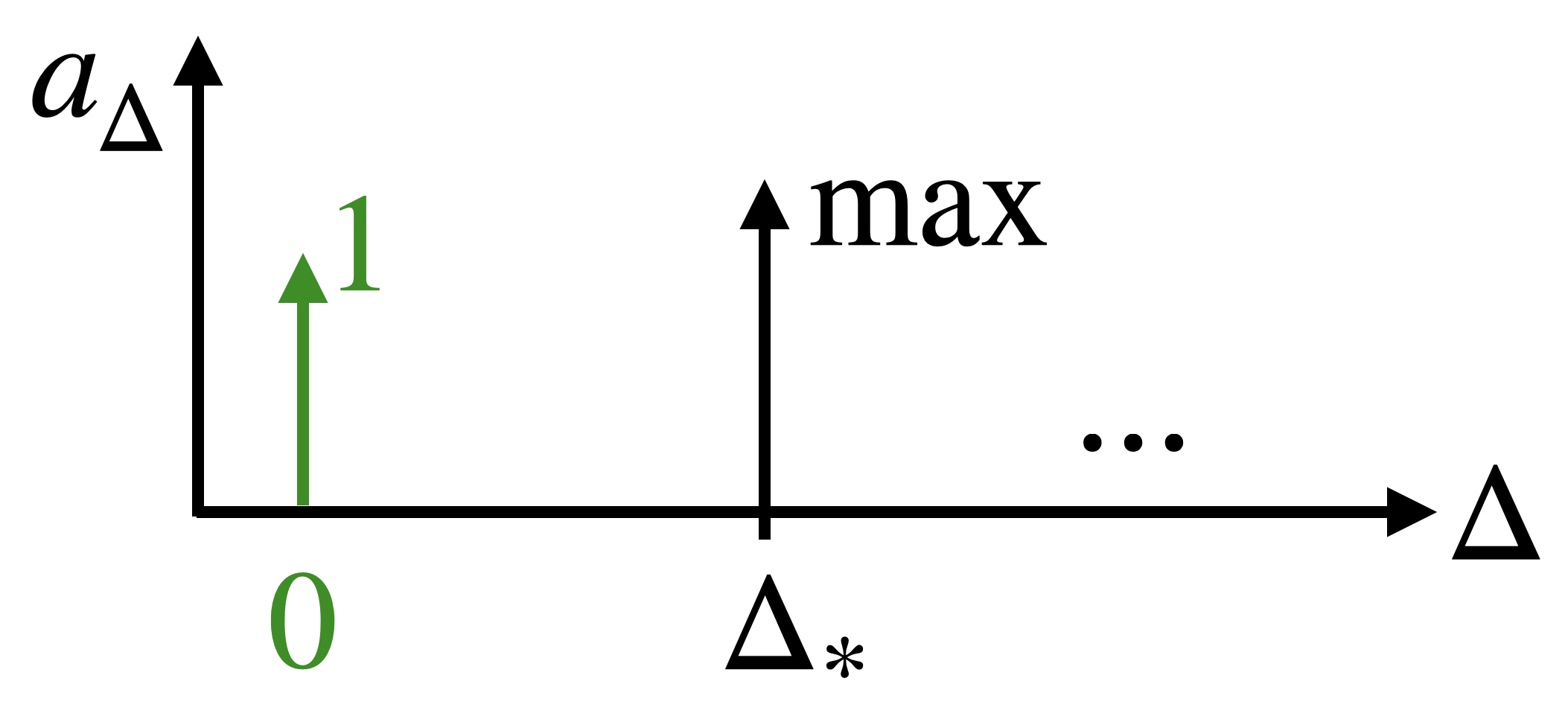}}
        \caption{A depiction of the optimisation problem. The source is in green.}
    \end{subfigure}\hspace{1cm}
    \begin{subfigure}[t]{0.53\textwidth}
        \centering
        \includegraphics[width=\linewidth]{phi4_flow.png}
        \caption{The extremal solution found numerically for this problem in \cite{Paulos:2019fkw} in the range $2\D_\phi\leq \Delta_*\leq 2\D_\phi+1$. Referred to as `$\phi^4$ flow', it interpolates between the GFB and GFF solutions. Reproduced from \cite{Paulos:2019fkw} with additional annotations.}
    \end{subfigure}
    \caption{The simplest OPE maximisation problem~\ref{prob:opemax} where the low-lying `source' only contains the identity operator, i.e. $S(z)=G_0(z)$.}
    \label{fig:phi4}
\end{figure}
\subsection{OPE maximisation}
The simplest version of OPE maximisation is to maximise the squared OPE coefficient $a_{\Delta_*}$ of the first operator $\Delta_*$ above the identity operator $\Delta=0$, subject to crossing and unitarity. For this to make sense, we need to impose a gap above the identity:
\be
\text{Gap assumption:} \quad  \text{The spectrum is contained in }\{0\}\cup[\Delta_*,\infty) \ ,
\ee
where the spectrum is the set $\Delta$ where $a_\Delta\neq 0$. This choice of gap is the input that specifies which theory we are looking for. As we will explain below, the problem turns out to be well-posed at least in the range $\Delta_* \in [\sqrt{2}\Delta_\phi, 2\Delta_\phi+1]$. 

\begin{problem}[OPE maximisation]\label{prob:opemax}
Fix the identity operator with $a_0=1$ and the gap to the next operator $\D_* \in [\sqrt{2}\Delta_\phi,2\Delta_\phi+1]$, so that the spectrum is contained in $\{0\}\cup [\Delta_*,\infty)$.
\begin{align}
  &\text{Maximise} &&  \hspace{-1cm}a_{\D_*} \no\\
  &\text{subject to} && \hspace{-1cm}\text{unitarity }\ a_\Delta\geq 0\no\\
  &&& \hspace{-1cm}\text{and crossing }\ \sum_\Delta a_\D\, G_\D(z)=\sum_\Delta a_\D\, G_\D(1-z)\no
\end{align}\label{OPEMaxProblem}
\end{problem}

Here $\Delta_\phi>0$ is arbitrary and fixed. As we reviewed in the introduction, in the range $2\Delta_\phi\leq \Delta_*\leq 2\Delta_\phi+1$, extremal solutions for this problem have been obtained numerically \cite{Paulos:2019fkw} and found to interpolate smoothly between the Generalised Free Boson (GFB) at $\Delta_*=2\D_\p$ with exchanged spectrum $2\D_\p+2n$, and the Generalised Free Fermion (GFF) at $\Delta_*=2\D_\p+1$ with exchanged spectrum $2\D_\p+2n+1$. We reproduce their plot here in Figure~\ref{fig:phi4}.

\begin{figure}[!t]
    \centering
    \begin{minipage}[t]{0.48\textwidth}
        \centering
        \includegraphics[width=2.3\dimexpr\paperwidth/7\relax]{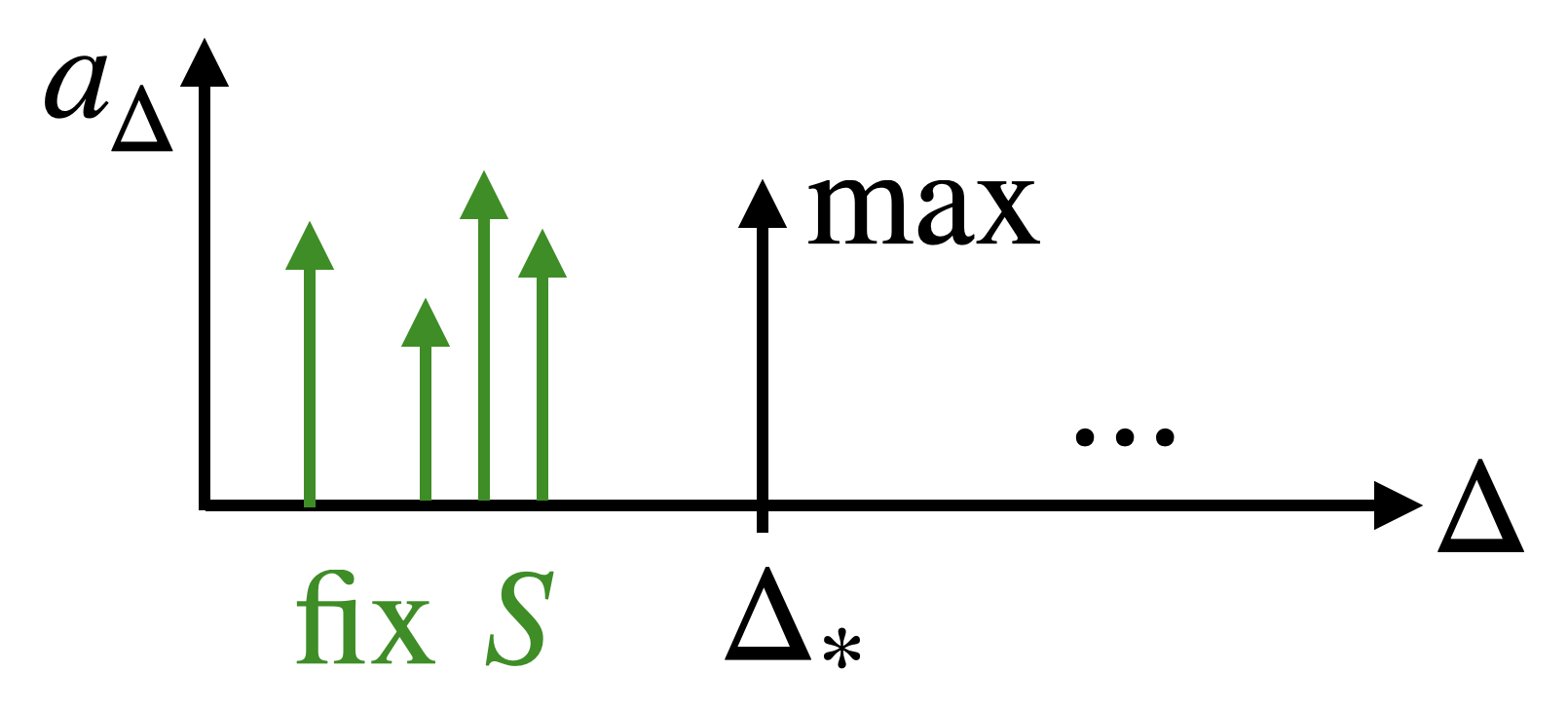}
        \caption{A depiction of a generalised OPE maximisation problem~\ref{prob:Gopemax}. The low-lying source is in green.}
                \label{fig:gOPE}
    \end{minipage}\hfill
    \begin{minipage}[t]{0.48\textwidth}
        \centering
\includegraphics[width=\dimexpr\paperwidth/16\relax]{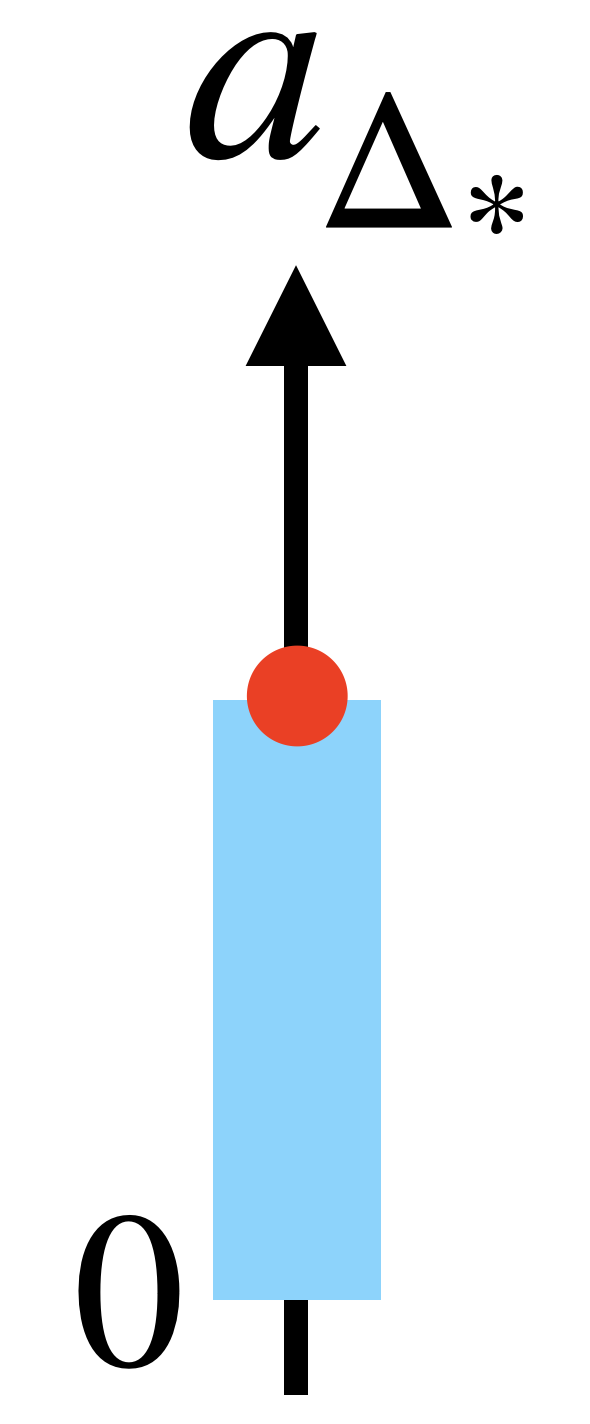}
        \caption{A sketch of the `islands' in the single variable $a_{\Delta_*}$ that we are maximising. This work proves that an extremal solution exists (shown in red), which attains the maximal value, if the problem is feasible.}
        \label{fig:island}
    \end{minipage}
\end{figure}

We will also consider the following generalised OPE maximisation problem, where the identity operator is replaced with an arbitrary `source' in the region $[0,\Delta_*)$ below the gap.
\begin{problem}[generalised OPE maximisation]\label{prob:Gopemax}
Fix any source $S(z)= \sum_{0\leq \Delta < \Delta_*} a_\Delta^{\rm fixed}\,  G_\Delta(z)$ below a gap $\D_* \geq \sqrt{2}\Delta_\phi$.
\begin{align}
  &\text{Maximise} &&  a_{\D_*} \no\\
  &\text{subject to} && \text{unitarity }\ a_\Delta\geq 0 \no \\
  &&&\text{and crossing }\ S(z)+\sum_{\Delta\geq \D_*} a_\D\, G_\D(z)=S(1-z)+\sum_{\Delta\geq \D_*}a_\D\, G_\D(1-z)\no
\end{align}
\end{problem}
\noindent Generalised OPE maximisation problems of this kind were discussed in \cite{Ghosh:2025sic}. See Figure~\ref{fig:gOPE} for a visual depiction of such a problem. We will generally allow the OPE density $a_\Delta$ to contain both discrete and continuous pieces. In that case, all OPE sums above should be interpreted as integrals. We shall explain this point further in Section \ref{sec:Convex-Hilbert}.

The allowed range of values of $a_{\Delta_*}$ can be seen as the simplest type of bootstrap island: an island in a single variable (see Figure \ref{fig:island}). We always have $a_{\Delta_*}\geq 0$ by unitarity, and the OPE maximisation problem asks what is the upper boundary of the island. 
This upper boundary is finite because of the assumption $\Delta_*\geq \sqrt{2}\D_\phi$, as we shall explain in the next subsection. The result of this work is the statement that, if the island contains anything at all, then it contains a point at the very maximum value of the island.

\subsection{Functionals and the dual problem}
The LHS of the crossing equation \eqref{crossblocks} is a function of the kinematic cross ratio $z$. The standard approach of the conformal bootstrap is to digest this kinematic dependence by acting with linear functionals. We shall introduce these concepts informally here, and they will be formally handled below in Section~\ref{subsec:1dCFTform}.

\paragraph{Functionals give sum rules.} We need to use an appropriate class of functionals $\omega$ that `swap' with the OPE sum \cite{Qiao:2017lkv}. Acting with $\omega$ on the crossing equation \eqref{crossblocks} then gives a sum rule
\be
\sum_\Delta a_\Delta \, \omega(\Delta) = \omega\left[ \sum_\Delta a_\Delta \, F_\Delta \right] = 0   \qquad \qquad \left(\omega(\Delta) \equiv \omega\left[F_\Delta\right] \right) \ . \label{sumrule}
\ee
This is true for any swappable functional $\omega$. In particular, $\omega$ could be enumerated over a complete basis for the appropriate class of functionals. We follow the standard convention of denoting by $\omega(\Delta)$ the action of the functional $\omega$ on the crossing vector $F_\Delta$.

A simple class of functionals are just evaluation at any point $y$ in the open disc $|y|<1$,
\be
e_y[F] := F(y) \ .
\ee
These satisfy swapping on unitary CFTs, which is simply the statement that the OPE converges absolutely at each $y$ in the open disc \cite{Pappadopulo:2012jk}. They are `complete' in the sense that if all of these vanish then $F=0$. These functionals can be rearranged by instead evaluating the derivatives at a point $|y|<1$,
\be
D^{(n)}_y[F] :=  F^{(n)}(y) \ , \qquad n=0,1,2,\ldots \ .
\ee
These similarly satisfy swapping by term-by-term differentiation of the OPE. For any fixed $y$, they are complete by analyticity: if all Taylor coefficients at $y$ vanish then $F=0$. It is this latter class of `derivative' functionals at the crossing symmetric point $y=0$ (or their analogue in the $z$ variable) that is typically used as a basis for the numerical conformal bootstrap.

\paragraph{Positive functionals imply bounds.}
Now suppose that a swappable functional $\omega$ has the following positivity properties:
\be
\omega(\Delta_*)\geq 1 \ , \qquad \omega(\Delta)\geq 0 \qquad \forall \Delta>\Delta_* \ , \label{posFunc}
\ee
i.e. its action on blocks is positive above the gap $\Delta_*$, and strictly positive at $\Delta_*$. We call such a functional a positive functional.

For such a functional, it is trivial to deduce that the sum rule \rf{sumrule} implies an upper bound on the OPE coefficient $a_{\Delta_*}$ that we are trying to maximise in Problem~\ref{prob:Gopemax}. The sum rule  samples the function $\omega(\Delta)$ against the weights $a_\Delta$. Since $a_\Delta\geq 0$ then the contribution from the tail $\Delta>\Delta_*$ is non-negative. Dropping this piece turns the equality \rf{sumrule} into an inequality,
\be
a_{\Delta_*} \leq - \frac{\omega[S^{(-)}]}{\omega(\Delta_*)} \ . \la{OPEbound}
\ee
This states that the positive contribution from $\Delta_*$ cannot be bigger than the negative contribution from the low-lying source $S(z)$. Here, and below, we denote the crossing-antisymmetric part of the source by
\be
S^{(-)}(z) = S(z) - S(1-z) \ .
\ee

For any positive functional $\omega$, the bound \rf{OPEbound} must be satisfied by any unitary solution of crossing. One can then optimize among positive functionals $\omega$ to make the bound as strong as possible. This is the `dual formulation' of the problem, which is the focus of the numerical conformal bootstrap:
\begin{equation}
      \begin{aligned}
      &\text{Minimise} && \qquad   -\omega[S^{(-)}] \\
  &\text{subject to} && \qquad  \text{positivity }\  \ \omega(\Delta_*)\geq 1 \ , \quad \omega(\Delta)\geq 0 \ \ \forall \Delta>\Delta_*  \ . \label{informalDual}
\end{aligned}
\end{equation}

Note that minimising $-\omega[S^{(-)}]$ is the same as minimising the RHS $- \frac{\omega[S^{(-)}]}{\omega(\Delta_*)}$ of the bound \rf{OPEbound}: because of the constraint $\omega(\Delta_*)\geq 1$, the best choice will always be to rescale the functional so that $\omega(\Delta_*)=1$.

As we will explain below, a version of this problem can be rigorously understood as the `convex dual' of the primal problem~\ref{prob:Gopemax} when formulated in an appropriate Hilbert space.

\paragraph{Boundedness of the primal problem.} 
In setting up the primal problems~\ref{prob:opemax} and~\ref{prob:Gopemax}, we assumed the gap is greater than a minimal value $\Delta_*\geq \sqrt{2}\D_\phi$. This guarantees that the problems are bounded, i.e.\ that the islands in which $a_{\Delta_*}$ is valued are bounded above. This is by a general mechanism: if the dual problem is \textit{feasible}, meaning there exists \textbf{some} functional with positivity properties \eqref{posFunc}, then it implies \textbf{some} bound on all solutions $a_{\Delta_*}$ of the primal problem \ref{prob:Gopemax}.

In our case, by Lemma~\ref{lemma:derivativePos}, if $\Delta_*\geq \sqrt{2}\D_\phi$ then the functional $\omega = D^{(1)}_0$ satisfies the positivity properties. Therefore, $\Delta_*\geq \sqrt{2}\D_\phi$ is a sufficient condition for boundedness of the primal problem.

Let us briefly analyse whether the condition $\Delta_*\geq \sqrt{2}\D_\phi$ can be relaxed while preserving boundedness. First, note that there is a hard minimum $\Delta_*>\frac43\D_\phi$. This follows because there are explicit, unitary solutions of crossing without identity with all values $0<\Delta_{\rm gap}\leq\frac43\D_\phi$ \cite{Antunes:2021abs,Cordova:2022pbl}. If $\Delta_*$ goes below this threshold, this solution can be added arbitrarily to primal solutions, making the objective function $a_{\Delta_*}$ unbounded. Therefore, the greatest possible relaxation of our assumption is mild ($\frac43=1.33...$ being close to $\sqrt{2}=1.41...$).

For any $\Delta_\phi$, our assumption can be slightly relaxed, since the functional $D^{(1)}$ is positive on a strictly larger range $(\Delta_{\rm pos},\infty)$, where $\frac43\D_\phi<\D_{\rm pos}<\sqrt{2}\Delta_\phi$. For small $\Delta_\phi \to 0$, one can show that $\D_{\rm pos}\to\frac43\D_\phi$ so one can relax $\Delta_*$ arbitrarily close to $\frac43\D_\phi$. In contrast, for large $\Delta_\phi \to \infty$, we find $\D_{\rm pos}\to\sqrt{2}\Delta_\phi$ so at least this functional does not allow any relaxation. While there may, of course, exist other positive functionals, we expect that our assumption becomes sharp in the $\Delta_\phi\to\infty$.\footnote{In the context of QFTs in AdS$_2$ and their boundary CFT$_1$ correlators, $\Delta_\phi\to\infty$ can be seen as a flat-space limit (with $R_{\rm AdS} \to\infty$ and $m^2=R^{-2}_{\rm AdS}\,  \D(\D-d)$ fixed) \cite{QFTinAdS,Mazac:2018mdx,Komatsu:2020sag,Cordova:2022pbl}. In the simplest case where the light source is just the identity block $S=G_0$, our assumption $\Delta_*\geq \sqrt{2}\D_\phi$ was conjectured to be sharp in this limit~\cite{Cordova:2022pbl}. In that case, the OPE-maximising `$\phi^4$ flow' CFT correlators become the following 2d integrable S-matrices in the flat-space limit: (i) the sine-Gordon lightest breather S-matrix (CDD pole) if $\sqrt{2}\Delta_\p\leq\D_*\leq2\D_\p$~\cite{Mazac:2018mdx} and (ii) the sinh-Gordon S-matrix (CDD zero) if $ 2\D_\p < \D_* \leq 2\Delta_\phi+1$ and $2\Delta_\phi+1-\Delta_* \sim \Delta_\phi^{-1/2}$ \cite{Cordova:2022pbl,Antunes:2024hrt}.}

\section{Conformal bootstrap as convex optimization}\label{sec:Convex-Hilbert}
In this section, we shall reformulate the OPE maximisation Problem~\ref{prob:Gopemax} as a convex optimisation problem in the Hilbert space $A^2$. By applying the Fenchel-Rockafellar theorem, we will prove primal attainment.

\subsection{Convex optimisation}\label{convOpt}
We will deal with a class of convex optimisation problems of the form 
\begin{align}
  \text{Minimise} \qquad  f(x)+g(x)  \ , \label{primalC}
\end{align}
where $x$ runs over a real Hilbert space $X$ and the \textit{objective functions} $f,g:X\to \mathbb{R}\cup\{+\infty\}$ are convex. If either $f$ or $g$ is infinite at some $x\in X$, that means the point $x$ is infeasible. We require $f$ and $g$ to be \textit{proper}, meaning neither of them is infinite everywhere.

We also require $f$ and $g$ to be \textit{lower semi-continuous} (lsc), meaning the sublevel sets $f^{-1}((-\infty,a])$ and $g^{-1}((-\infty,a])$ are closed for all $a\in \mathbb{R}$. To give some illustration of what lsc means, note that continuous functions are lsc. Conversely, if $f$ and $-f$ are both lsc, then $f$ is continuous. Other examples of lsc functions include \textit{indicator functions} on closed sets $K\subset X$,
\be
1_K(x):=
\begin{cases}
0 &,  \qquad x\in K\ ,\\
+ \infty &, \qquad  x\notin K\ .
\end{cases}
\ee
The indicator function is also convex if the set $K$ is convex. Therefore, by adding indicator functions to $f$ or $g$, we can impose  closed, convex constraints: demanding that $x\in K$.

\paragraph{Primal and dual problems.} Let us call \rf{primalC} the primal problem, and denote its optimal value by 
\be
P := \inf_{x\in X}(f(x)+g(x))~. \label{PrimalProblem}
\ee
Note that this minimum need not be attained. For an infeasible problem, we would have $P=+\infty$. For an unbounded problem, we would have $P=-\infty$.

Each convex optimisation problem of this form has \textit{dual problem}, which is formulated in terms of \textit{convex conjugate} functions $f^*,g^*:X \to\mathbb{R}\cup\{+\infty\}$ defined by
\begin{align}
    &f^*(y) = \sup_{x\in X}\big(\langle y|x\rangle -f(x)\big) \ , \label{FenchelDual}
\end{align}
and similarly for $g$. The dual problem is then to maximise $-f^*(y)-g^*(-y)$:
\begin{equation}
    D:=\sup_{y\in X}\big(-f^*(y)-g^*(-y)\big)\ .\label{DualProblem}
\end{equation}
The split into $f$ and $g$ in the primal problem may have seemed arbitrary: we might as well have defined $h=f+g$ and asked to minimise $h$. However, the dual problem actually depends non-trivially on the choice of $f$ and $g$. Moreover, the property we will study below---a regularity condition for the dual problem---will depend importantly on this choice.

Note that if $f$ and $g$ are proper, convex and lsc, then so are $f^*$ and $g^*$. The Fenchel-Moreau theorem states that their convex conjugates coincide with the original functions:
\be
f^{**}=f \ .
\ee
We can therefore regard the dual as a convex optimisation problem in its own right, where we have replaced $f(x) \to f^*(x)$ and $g(x) \to g^*(-x)$. Treating that instead as the primal problem, then its dual is $P$. The whole setup is therefore invariant under swapping the primal and dual problems.

\paragraph{Duality.}
To see why we call $D$ the dual of $P$, suppose $x\in X$ is any feasible solution of $P$ and $y\in X$ is any feasible solution of $D$. Feasibility simply means $f(x)+g(x)$ and $-f^*(y)-g^*(-y)$ are finite. By definition of the convex conjugate functions in terms of the supremum, we trivially have
\be
f^*(y) \geq \langle y|x\rangle - f(x) \ , \qquad \qquad  g^*(-y) \geq -\langle y|x\rangle -g(x) \ .
\ee
The sum of these inequalities gives
\be
- f^*(y) -g^*(-y) \leq f(x) + g(x) \ . \label{PDineq}
\ee
This means that all primal feasible values are greater or equal to all dual feasible values. As in the informal bootstrap discussion above, we can therefore interpret elements $y\in X\cong X'$ as functionals associated to bounds: all $x\in X$ must satisfy the lower bound \rf{PDineq}.

Taking the supremum of \rf{PDineq} over $y\in X$ and the infimum over $x\in X$, we trivially obtain the following statement:
\begin{proposition}[\textit{Weak duality}]
The primal and dual optimal values satisfy the inequality
\be
D\leq P \ .\label{WeakDuality}
\ee
\end{proposition}
\noindent In particular, if both problems are feasible then this implies that $P$ and $D$ are both finite: neither problem can be unbounded.

The power of this framework is that, under additional regularity assumptions, these primal and dual optima are not only ordered in this way, but they are actually equal. This is called \textit{strong duality}:
\begin{theorem}[Fenchel-Rockafellar~\cite{Rockafellar66,Rockafellar74}]\label{thm:FR}
    Let $X$ be a real Hilbert space and $f,g:X\rightarrow \mathbb{R}\cup\{+ \infty\}$ be proper, convex and lsc. Then:
    \begin{enumerate}[(i)]
        \item If the primal problem satisfies the regularity condition
        \begin{equation}
            0\in \operatorname{int}(\operatorname{dom} f - \operatorname{dom}g), \label{GSlater1}
        \end{equation}
        then $P=D$ and the dual optimum $D$ is attained if it is finite.
        \item If the dual problem satisfies the regularity condition
        \begin{equation}
            0\in \operatorname{int}(\operatorname{dom} f^* + \operatorname{dom}g^*), \label{GSlater2}
        \end{equation}
        then $P=D$ and the primal optimum $P$ is attained if it is finite.
    \end{enumerate}
\end{theorem}
\noindent The second part of the theorem follows trivially from the first part by swapping $P$ and $D$. We state it explicitly since we shall use the second part below.

The regularity conditions \rf{GSlater1} and \rf{GSlater2} can be seen as infinite-dimensional generalisations of Slater's condition. We found \cite{boct2011comparison} a useful resource to learn about the available regularity conditions. Note that the Hilbert space assumption is not strictly necessary for this theorem, but is a simplifying property and is satisfied in our case.

\subsection{Formalism for 1d CFT}\label{subsec:1dCFTform}

In order to apply the Fenchel-Rockafellar Theorem \ref{thm:FR} to OPE-max, we need to formulate that problem in the appropriate formalism. We shall set out the formulation of OPE coefficients as Radon measures, and of correlation functions as elements of a Hilbert space.

\paragraph{OPE coefficients as Radon measures.} The physical solutions of most interest typically have discrete spectrum. However, the most general possibility is that the OPE coefficients $a_\Delta$ may contain both discrete and continuous contributions. Indeed, 1d CFTs with continuous spectrum do exist, for example the subsector of 2d Liouville CFT  with all operators placed on a common circle.

In order to apply our functional analytic machinery, we must remain in this most general case. Formally, we take $a_\Delta$ to be a \textit{Radon measure} in the variable $\Delta$. This is defined as a locally integrable (Borel) measure.\footnote{On more general spaces, there is an additional regularity assumption for a Borel measure to be Radon, but this is automatic on certain spaces, including locally compact, second countable Hausdorff spaces.} The condition of local integrability is required for the OPE sum \rf{OPE} to converge at least on compact regions $\Delta\leq N$, i.e.\ forbidding accumulation points where the OPE fails to converge.  Radon measures can also be characterised as the continuous dual of the space of compactly supported continuous functions, meaning they can naturally be integrated against such functions.

A key point is that examples of Radon measures include discrete atoms and continuous densities, e.g.
\be
a(\Delta) = \sum_i a_{\Delta_i}\delta_{\Delta_i} + \a(\D)\, \ud\D    \quad \to  \quad G(z) = \sum_i a_{\Delta_i} G_{\Delta_i}(z) + \int\ud\D  \,  \a(\D)\, G_{\Delta}(z)  , \la{dens}
\ee
where $\delta_\Delta$ denotes an atom at $\Delta$ with the property $\int_S \delta_\Delta=1$ if $\Delta\in S$. 

Let us therefore clarify how the OPE-max problems \ref{prob:opemax} and \ref{prob:Gopemax} are interpreted in the case where $a(\Delta)$ is a Radon measure:
\begin{itemize}
    \item As mentioned above, all OPE sums should be interpreted as integrals over $\Delta$. For discrete contributions, these reduce again to the usual sums.
    \item The objective function 
    \be
    a_{\Delta_*} \equiv a_{\{\Delta_*\}} \equiv \lim_{\epsilon\to 0} \int_{{\Delta_*}-\epsilon}^{{\Delta_*}+\epsilon} \ud a(\Delta) 
    \ee
    is the \textit{discrete contribution} at $\Delta_*$, i.e.\ the coefficient of the atom $\delta_{\Delta_*}$. It receives no contribution from continuous pieces of $a(\Delta)$.\footnote{Note that this interpretation is unavoidable, unless one instead maximises the weight on a finite interval $\int_{[\Delta,\Delta_*+\epsilon]}\ud a(\Delta)$, but that fails to be physically interpretable as `maximising the OPE coefficient at $\Delta_*$'.}
    
    \item The unitarity condition ``$a_\Delta\geq 0$'' is defined by 
    \be
\int \ud a(\Delta) \, f(\Delta) \geq 0
    \ee
    for all $f\geq 0$ continuous and compactly supported.
\end{itemize}

For clarity, let us state the OPE maximisation problem formally in terms of Radon measures:
\begin{problem}[OPE-max with Radon measures]\label{prob:Fopemax}
Fix a gap $\D_* \geq \sqrt{2}\Delta_\phi$ and a source $S(z)= \int_{[0,\D_*)} \ud a^{\rm fixed}(\Delta)\,  G_\Delta(z)$, where $a^{\rm fixed}(\Delta)$ is a  finite positive Radon measure on $[0,\Delta_*)$, i.e.\ strictly below a gap. Let $a(\Delta)$ be a Radon measure on $[\Delta_*,\infty)$.
\begin{align}
  &\text{Maximise} &&  a(\{\Delta_*\}) \no\\
  &\text{subject to} && \text{unitarity }  \ \int_{[\Delta_*,\infty)} \ud a(\Delta)\, f(\Delta) \geq 0 \ \text{ for }f\geq 0\text{ continuous with compact support}\no \\
  &&&\text{and crossing }\  \ S(z)+\int_{[\D_*,\infty)} \ud a(\D)\, G_\D(z)=S(1-z)+\int_{[\D_*,\infty)} \ud a(\D)\, G_\D(1-z) \ . \no 
\end{align}
The statement of crossing applies at each $z\in \mathbb{C}\setminus\left((-\infty,0]\cup[1,\infty)\right)$, and demands that the OPE sum $\int_{[\D_*,\infty)} a_\D\, G_\D(z)$ converges at each point.
\end{problem}
\noindent This is simply how Problem~\ref{prob:Gopemax} is interpreted in terms of Radon measures, which allow for both discrete and continuous spectra.

\paragraph{Bergman space of correlators.}  There is some freedom over which space is used to model correlation functions. We find the following \textit{Bergman space} to be a natural choice
\be
A^2 := \left\{ G(y) \text{ analytic on }\mathbb{D} \ \middle| \ \int_\mathbb{D} \ud^2 y \ w(y) \, |G(y)|^2 < \infty  \right\} \ , \label{A2def}
\ee
with weight function
\be
w(y) = \frac{\left(1-|y|^2\right)^\gamma}{|1-y^2|^\alpha\, |1+y^2|^\beta} \ ,
\ee
depending on 3 real parameters $\a,\b,\gamma$. The parameter $\gamma$ controls how hard elements are allowed to behave at generic points on the boundary of the disc, while $\alpha$ and $\beta$ allow specific control at the points $y=\pm 1$ and $y=\pm i$ respectively. Recall that $y$ is the Zhukovsky variable \eqref{eq:crossrel}, and those special points map to the branch points $z=0,1$ and $z=\infty$.

As we will show below, unitary correlators $G(y)$ belong to this space if\kern.1em\footnote{We will abuse notation by writing $G(y) \equiv G(z(y))$, etc., when we deem the meaning to be clear.}${}^{,}$\footnote{In fact, one could decrease $\gamma$ slightly further to $\gamma>\max\{8\Delta_\phi-2,-1\}$ with correlators still belonging to the space. We believe that all of our results remain true in that case.} 
\be
\gamma > 8\Delta_\phi -1 \ , \qquad \qquad \a<\gamma+2-8\Delta_\phi \ , \qquad \b<\gamma+2 \ . \label{eq:gamma-range}
\ee
It is trivial to verify that individual blocks $G_\Delta(y)$ with $\Delta\geq 0$ belong to this space under these conditions. Our primal attainment results and the rest of the paper will not depend on these parameters, so one could just set $\alpha=\beta=0$ for simplicity. However, for generality, for the relation to other works, and for future studies of dual attainment, we prefer to retain this most general case.

$A^2$ is a separable Hilbert space over the reals with the inner product
\be
\langle G\, | \, H\rangle = \Re \int_\mathbb{D} \ud^2 y \  w(y)\, \overline{G(y)} \, H(y) \label{eq:IP}
\ee
as long as $\gamma>-1$, which follows from \eqref{eq:gamma-range} and $\Delta_\phi>0$. It will be highly convenient below to deal with a Hilbert space, rather than wilder Banach or Fr\'echet spaces, for one thing because Hilbert spaces are automatically reflexive---that is, the canonical embedding $\mathcal H \hookrightarrow\mathcal H^{**}$ is an isomorphism.

As a first sense check, one can easily show using Lemma~\ref{lemma:RadDom} that the maps $\Delta \mapsto G_\Delta$ and  $\Delta \mapsto F_\Delta$ from $[0,\infty) \to A^2$ are continuous with respect to the Bergman norm. The following lemma further shows that the Bergman norm is a natural one for conformal field theory.
\begin{lemma}\label{lemma:OPEA2}
    Let $G =\int \ud a(\Delta) \, G_\Delta$ be a unitary solution of crossing, where $a(\Delta)$ is a Radon measure on $[0,\infty)$. Then $G$  belongs to $A^2$ and the OPE converges in the Bergman norm,
    \be
G = \lim_{N\to \infty} \int_0^N \ud a(\Delta) \, G_\Delta {\rm\ \  in \  }A^2 \ . \label{OPEconvB}
    \ee
    In particular, the statement of crossing is Bergman-convergent,
    \be
 \lim_{N\to \infty} \int_0^N \ud a(\Delta) \, F_\Delta = 0 \ . \label{crossdec}
    \ee
\end{lemma}
\begin{proof} 

Our strategy is to derive a uniform bound $\int \ud a(\Delta) \, |G_\Delta (y)| \leq M(y)$ on the absolute OPE sum, where the function $M(y)$ should have finite Bergman norm (even if it is not analytic, so not formally an element of the space). This will then give us control over the norm convergence of the OPE.

First note that in the $\rho$ variable, blocks satisfy the well-known inequality
\be
|G_\Delta(\rho)| \leq \left|\frac{z(|\rho|)}{z(\rho)}\right|^{2\Delta_\phi } G_\Delta(|\rho|) \ .
\ee
Summing against the positive OPE measure gives
\be
\int \ud a(\Delta) \, |G_\Delta (\rho)| \leq \left|\frac{z(|\rho|)}{z(\rho)}\right|^{2\Delta_\phi }   \int \ud a(\Delta) \, G_\Delta (|\rho|) = \left|\frac{z(|\rho|)}{z(\rho)}\right|^{2\Delta_\phi }   G(|\rho|) \ . \label{eq:GGa}
\ee
Now, how are correlators $G(|\rho|)$ satisfying crossing and unitarity allowed to behave on physical interval $|\rho|\in[0,1)$, or equivalently $|z|\in[0,1)$? Working in the $z$ variable, it can certainly blow up as $|z|^{-2\Delta_\phi}$ at $z=0$, since the definition 
\rf{OPE} includes this factor---and the same at the point $z=1$ related by crossing. We shall show that this is the worst possible behaviour. The function $|z|^{2\Delta_\phi}G(|z|)=\int \ud a(\Delta) \, |z|^{\Delta}\, {}_2F_1(\Delta,\D;2\D;|z|)$ is increasing over $|z|\in[0,1)$, so on $|z|\in(0,1/2]$, we have $G(|z|)\leq |2z|^{-2\Delta_\phi}G(\ha)$. By crossing, we also get $G(|z|)\leq (2|1-z|)^{-2\Delta_\phi}G(\ha)$ on $[\ha,1)$. Since both $ |z|^{-2\Delta_\phi}$ and $|1-z|^{-2\Delta_\phi}$ are bounded by $|z|^{-2\Delta_\phi}|1-z|^{-2\Delta_\phi}$, we conclude that $G(|z|)\leq C\, |z|^{-2\Delta_\phi}|1-z|^{-2\Delta_\phi}$ on all of $(0,1)$, for a constant $C$. Applying this to the RHS of \rf{eq:GGa} gives
\be
\int \ud a(\Delta) \, |G_\Delta (\rho)| \leq M(\rho) := C \ |z(\rho)|^{-2\Delta_\phi}\,  |1-z(|\rho|)|^{-2\Delta_\phi} \ . \la{majorant}
\ee
The RHS only blows up at $\rho=0$ and $|\rho|=1$, i.e.\ at $z=0$ and the cut $z\in[1,\infty)$. But one can check that  it is $O(1)$ when approaching $\rho=-1$ (i.e.\ $z=\infty$) along any direction that is not parallel to the cut. Equivalently in the $y$ variable, it diverges as $|y+1|^{-4\Delta_\phi}$ at $-1$ and as $|1+y^2|^{4\Delta_\phi
}(1-|y|)^{-4\D_\phi}$ at the arcs of the unit circle between $i \to 1$ and $1 \to -i$ (remaining finite when approaching the endpoints $\pm i$ non-tangentially to the circle). Hence, one finds that the Bergman norm $\|M\|^2=\langle M| M\rangle$ of $M(y)$ is finite precisely if the conditions \eqref{eq:gamma-range} are satisfied.

It follows that the correlator $G(y)$ belongs to $A^2$. To see this, note that it is holomorphic on the disc because the OPE converges locally uniformly, as follows from radial dominance (Lemma~\ref{lemma:RadDom}). Moreover, by \rf{majorant}, its Bergman norm is less than that of $M(y)$, so is finite.

The convergence of the OPE is dominated, by \rf{majorant}, as
\be
\left|\int_0^N \ud a(\Delta) \, G_\Delta(y) \right|\leq \int_0^N \ud a(\Delta) \, |G_\Delta (y)|  \leq M(y) \ .
\ee
Hence by Dominated Convergence, we obtain
\be
\left\| G - \int_0^N \ud a(\Delta) G_\Delta \right\|^2 \to 0 \ ,
\ee
as required for \rf{OPEconvB}.

For \rf{crossdec}, we can replace $z\mapsto1-z$ in all of the above. By crossing, we have $G(z)-G(1-z)=0$.  Decomposing with the OPE \rf{OPEconvB} for each term, we indeed find that the integral of crossing vectors converges to zero in Bergman norm.
\end{proof}

The above argument is essentially equivalent to a Tauberian one \cite{Qiao:2017xif}: the identity exchange in the crossed channel controls how $a(\Delta)$ behaves at large $\Delta$, and this forces convergence of the OPE in a uniform manner. A similar argument was used in \cite{Kravchuk:2020scc} to prove convergence of the OPE as a tempered distribution (the boundary value of the analytic functions on the unit circle). Bergman convergence is actually stronger, and implies that tempered distributional convergence.

\paragraph{Bergman functionals.} 
Now that we have set up the space of correlators, there is a natural class of dual functionals. By the Riesz representation theorem,  a Hilbert space is canonically isomorphic to its space of continuous linear functionals:
\be
A^2 \cong (A^2)' \ .
\ee
In other words, any element $h \in A^2$ has a continuous linear action $\omega_h:A^2\to \mathbb{R}$
\be
\omega_h[F] = \langle h|F\rangle \ ,
\ee
and all continuous functionals take this form. We shall call these \textit{Bergman functionals}. The following lemma shows that they are well behaved and have all of the properties that we want.

\begin{lemma}
    When acting on unitary solutions of crossing, Bergman functionals swap with the OPE and give absolutely convergent sum rules
    \be
\int_0^\infty \ud a(\Delta)\,  \omega_h(\Delta) = 0 \ , \label{BergmanSR}
    \ee
    where, as above, we denote $\omega_h(\Delta) = \omega_h[F_\D]=\Re\, \langle h \, |\, F_\D\rangle$.
\end{lemma}
\begin{proof}
    The sum rule will follow by taking the inner product of \rf{crossdec} against $h$. To prove absolute convergence, consider the resulting double integral, with absolute value of the integrand:
    \be
\int_{\mathbb{D}} \ud^2 y\, \int_0^\infty \ud a(\D)\,  w(y) \, \left| h(y) \, F_\Delta(y) \right| =   \int_{\mathbb{D}} \ud^2 y\, w(y) \, \left| h(y)\right| \  \int_0^\infty \ud a(\D)\,\left| F_\Delta(y) \right|
    \ee
    The bound \eqref{majorant} implies that
\begin{align}
\int_{\mathbb{D}} \ud^2 y\, \int_0^\infty \ud a(\D)\,  w(y) \, \left| h(y) \, F_\Delta(y) \right| &\leq \int_{\mathbb{D}} \ud^2 y\,   w(y) \, |h(y)| \, (M(y)+M(-y)) < \infty \ . \label{AbsConv}
\end{align}
Finiteness of this integral follows by the Cauchy-Schwarz inequality and the fact that $h(y)$, $M(y)$, and hence $M(-y)$ have finite Bergman norms.

Armed with this absolute convergence, we can simply take the inner product of \rf{crossdec} with $h$, exchange the order of the inner-product with the limit and integral, and obtain the sum rule \rf{BergmanSR}. It converges absolutely by comparison with \rf{AbsConv}.
\end{proof}

The Bergman space is a very large class of functionals. It contains the real and imaginary parts of the evaluation functional at each point $y\in \mathbb{D}$,\footnote{The continuous functionals here are the real and imaginary parts, since we are treating $A^2$ as a real Hilbert space---hence the factors of $i$ in \rf{eq:eval}.}
\be
e_y [F] := F(y) = \langle R_y | F \rangle + i\,  \langle i R_y | F \rangle \, \label{eq:eval}
\ee
which is given by the so-called \textit{reproducing kernel},
\be
R_y(u) = \frac{\gamma+1}{\pi} \,  \frac{\chi(u)\, \overline{\chi(y)}}{(1-u\overline y)^{\gamma+2}} \  , \quad \qquad \chi(y)=(1-y^2)^{\alpha/2}(1+y^2)^{\beta/2} \ . \label{repker}
\ee
The derivative functionals of all orders at each point $y$ also belong to this class. They are obtained by simply taking derivatives of the reproducing kernel
\be
D^{(n)}_y [F] := F^{(n)}(y) = \langle (\partial_{\bar y})^nR_y | F \rangle
 + i \, \langle i  (\partial_{\bar y})^nR_y | F \rangle
\ee
At any fixed $y\in \mathbb{D}$, the derivative functionals $D_y^{(n)}$ with $n\geq 0$ are a complete, minimal set for the space $A^2$. This follows by simply Taylor expanding the function $F(y)$. Their completeness is an attractive feature of the Bergman space, and it matches the prior observation that analytic functional kernels have convergent expansions  into monomials in the Zhukovsky variable~\cite{Mazac:2016qev}.

Furthermore, the class of Bergman functionals contains all of the general functionals considered by Maz\'a\v{c} and Paulos \cite{Mazac:2018mdx} and Levine and Paulos \cite{part2}, for suitable choices of the parameters satisfying \eqref{eq:gamma-range}. Those functionals were given by contour integrals against analytic functional kernels $f(y)$,
\be
\omega_f[F] = \frac{1}{2\pi i} \int_C \ud y\, f(y) \, F(y)
\ee
along a prescribed contour $C$. The embedding of these functionals into $A^2$ can be written explicitly in terms of the reproducing kernel  \rf{repker}:\footnote{For example, consider the simple kernel $\tilde f_0(z)=z^{-2}$ in the $z$ variable, which is the first `single zero' functional $\tau_1$ studied in Appendix D of \cite{Ghosh:2025sic}. In the $y$ variable, after including the Jacobian factor, it translates to $f_0(y)=4(1-y)/(1+y)^3$. Making the simple choice for the space $(\a,\b,\g)=(0,2,1)$ (which requires $\Delta_\phi<\tfrac 14$ for \eqref{eq:gamma-range}), we obtain the Bergman element
\be
G_{f_0}(y) =\frac{8\,\big(8y^3-5y^2+8y-3\big)}{\pi^2\,(1+y)^4}
+\frac{16\,\big(1+y^2\big)\big(5y^2-5y+2\big)}{\pi^2\,(1+y)^5}\, \arctan \frac{1+y}{1-y}\ .
\ee}
\be
G_f(u) = -\frac{1}{2\pi i} \int_{C} d\overline y \ \overline{f(y)}\,R_{y}(u) \quad \in\quad  A^2 \ .
\ee

In the argument below we will use the existence of the positive functional constructed in the following lemma.
\begin{lemma}\label{lemma:posFunc}
    For any $y_0\in(0,1)$, there exists a functional $h_{y_0}\in A^2$ with the following properties:
    \begin{enumerate}[(i)]
        \item Strict positivity $h_{y_0}(\Delta) >0$ for $\Delta\geq \sqrt{2}\Delta_\phi$,
        \item Dominance over blocks and crossing vectors on a disc of radius $y_0$,
        \be
\frac{G_\Delta(y)}{h_{y_0}(\Delta)} \overset{\Delta\to\infty}\to 0  , \qquad \frac{F_\Delta(y)}{h_{y_0}(\Delta)} \overset{\Delta\to\infty}\to 0  \qquad \quad \forall\  |y|<y_0 \ .
        \ee
    \end{enumerate}
\end{lemma}
\begin{proof}First consider the functional $D_{0}^{(1)}$ that evaluates the first derivative at $y=0$. Its action $D^{(1)}_{0}(\Delta)\equiv D^{(1)}_{0}[F_\Delta]$ on crossing vectors is real and strictly positive on $\Delta\geq \sqrt{2}\Delta_\phi$ by Lemma~\ref{lemma:derivativePos}. Since $\D \mapsto F_\Delta$ is continuous, then $D^{(1)}_{0}(\Delta)$ is bounded away from zero on the compact interval $[\sqrt{2}\Delta_\phi,2\Delta_\phi]$.

    Now let us define 
    \be
h_{y_0} = \Re\big[ D^{(1)}_{0} + \alpha_{y_0} e_{y_0} \big] \ .
    \ee
    We only know that $e_{y_0}(\D)$ real and is positive on $[2\Delta_\phi,\infty)$ but it is bounded on $[\sqrt{2}\Delta_\phi,2\Delta_\phi]$ since it is continuous. Therefore, for sufficiently small $\alpha_{y_0}>0$, the above combination is positive on the whole region $[\sqrt{2}\Delta_\phi,\infty)$.

    For dominance, at large $\Delta$ the denominator satisfies $h_{y_0}(\Delta)\geq \alpha_{y_0} F_{\Delta}(y_0)$. By Lemma~\ref{lemma:crossPos}, we then get
    \be
    h_{y_0}(\Delta)\geq \alpha_{y_0}(1-\kappa_{z(y_0)})\, G_{\Delta}(y_0) \ . \label{eq:interEq}
    \ee
    It is sufficient to prove the dominance over blocks, since crossing vectors satisfy $|F_\Delta(y)|\leq |G_\Delta(y)|+|G_\Delta(-y)|$. Using \eqref{eq:interEq} and  Lemma~\ref{lemma:RadDom}, we find
    \be
\left|\frac{G_\Delta(y)}{h_{y_0}(\Delta)}\right| \leq C_{y,y_0}\, \frac{G_\Delta(|y|)}{G_\Delta(y_0)} \ ,
    \ee
    where $C_{y,y_0}$ does not depend on $\Delta$. This quantity decays exponentially at large $\Delta$ by Lemma~\ref{lemma:ratBlocks}.
\end{proof}

\subsection{Formulating OPE maximisation in Bergman space}
We shall now formulate an optimisation problem in $A^2$ which is equivalent to the physical OPE-max problem.

Let us focus on the odd part of the correlator, which is supposed to vanish by crossing,
\be
F(z) = \int_\Delta \ud a(\Delta) \, F_\Delta(z) \in A^2 \  , \qquad F_\Delta(z) = G_\Delta(z) - G_\Delta(1-z) \ .
\ee
We would schematically like to take
\be
f(F) = a_{\Delta_*}  + 1_{\{a_\Delta\geq 0\}} \ , \qquad g(F) = 1_{\{F=0\} } \ .
\ee
Here $g$ imposes the crossing constraint $F=0$ and is straightforward. On the other hand, $f$ contains the objective function (maximise $a_{\Delta_*}$, or rather minimise $-a_{\Delta_*}$) and imposes the positivity constraint. But as it stands, this definition of $f(F)$ does not make sense, since the quantities $a_\Delta$ are not well-defined as functions of $F$.

\paragraph{Positivity condition.}
First, let us understand how to impose positivity. Define the \textit{positive cone} by
\be
C_+ := \left\{ \lim_{N\to\infty} \int_{[\Delta_*,N)}  \ud a(\Delta)\,  F_\Delta \ \middle| \  a(\Delta) \text{ a non-negative Radon measure, }\lim_{N\to\infty}\text{ exists in }A^2  \right\} \ .
\ee
Since we need the constraint set to be closed, we shall take its closure $\overline{C_+}$ in $A^2$, which we shall call the \textit{closed positive cone}. Because of the following lemma, taking this closure is innocuous and will have no effect on the primal attainment argument.\footnote{Incidentally, we believe that $C_+$ is already closed at least in the special case $\alpha\geq |\beta|$. In that case, it follows from the positivity of the expansion of crossing vectors in the Zhukovsky variable $y$ around $y=0$, i.e.\ $F_\Delta(y) = \sum_n b_n(\Delta) y^n$ with $b_n(\Delta)\geq0$ for $\Delta>C$ with $C$ independent of $n$. This expansion is not directly related to an OPE expansion, and we are not aware of any previous statement of its positivity. We conjecture it to be true based on ample numerical evidence, but currently no analytic proof. In any case, this plays no role in our argument.} 
\begin{lemma}\label{lemma:clcone}
    Every element of the closed positive cone $\overline{C_+}$ takes the form
    \be
F(y) = \int_{[\Delta_*,\infty)} \ud a(\Delta)\,F_\Delta(y)
    \ee
    for $a(\Delta)$ a non-negative Radon measure, and with the integral converging absolutely at each $y \in \mathbb{D}$. The OPE sum $\int_{[\Delta_*,\infty)} \ud a(\Delta)\,G_\Delta(y)$ also converges absolutely at each $y \in \mathbb{D}$.
\end{lemma}
\begin{proof}
    Suppose $F = \lim_{n\to \infty}F^{(n)}\in \overline{C_+}$ is an $A^2$-limit of elements
    \be
F^{(n)}(y) =  \int_{[\Delta_*,\infty)} \ud a^{(n)}(\Delta)\,F_\Delta(y) \in C_+
    \ee
First note that $F(y)$ is analytic in $\mathbb{D}$ because Bergman convergence implies locally uniform convergence (this is why the space is closed: Bergman limits preserve analyticity).
    
    Fix any $y_0\in(0,1)$ and define Radon measures rescaled by the functional action of Lemma~\ref{lemma:posFunc}
    \be
\nu^{(n)}(\Delta) = h_{y_0}(\D) \, a^{(n)}(\Delta) \ .
    \ee
    These are positive since $h_{y_0}(\D)\geq0$ for $\Delta \geq \Delta_*\geq \sqrt{2}\Delta_\phi$.

    Their total masses are a convergent positive sequence so they are bounded:
\be
\int |\nu^{(n)}|= \int \ud a^{(n)}(\Delta)\, h_{y_0}[F_\Delta] = h_{y_0}[F^{(n)}] \to h_{y_0}[F]
\ee
where we have used the fact that $h_{y_0}\in A^2$ defines a continuous functional.

 Now we invoke an important property of Radon measures: \textit{Any sequence of Radon measures with uniformly bounded mass has a convergent subsequence.}\footnote{This follows from the Banach-Alaoglu theorem applied to the space of continuous functions on $[\Delta_*,\infty)$ vanishing at $\infty$, whose continuous dual is the space of Radon measures.} Convergence here is in the `vague', i.e.\ weak$^*$ topology. Thus, after passing to a subsequence, we have $\nu^{(n)} \to \nu$ for some positive Radon measure $\nu(\Delta)$ on $\Delta\in[\Delta_*,\infty)$. 
 
Define a positive Radon measure
\be
a(\Delta) := h_{y_0}(\D)^{-1} \, \nu(\D) \ .
\ee
For any $|y|<y_0$, by Lemma~\ref{lemma:posFunc}, we have $|F_\D(y)/h_{y_0}(\D)| \to 0$ as $\D\to\infty$. Since integration against a continuous function vanishing at $\Delta=\infty$ is a bounded functional on Radon measures, we obtain
\be
F(y) = \lim_{n\to\infty} F^{(n)}(y) = \lim_{n\to\infty} \int \frac{F_\D(y)}{h_{y_0}(\D)}\, \ud\nu^{(n)}(\Delta) = \int \frac{F_\D(y)}{h_{y_0}(\D)}\, \ud\nu(\Delta) = \int \ud a(\Delta) \, F_\Delta(y) \ . \la{longeq}
\ee

So far we have proved the claim for $|y|<y_0$. But $y_0$ was arbitrary: the final step is to prove that $a(\Delta)$ does not depend on $y_0\in(\ha,1)$. Consider a larger value $y_0'\in (y_0,1)$. After passing to a further subsequence, we similarly find that 
\be
\nu^{(n)}{}'(\Delta) := \frac{h_{y_0'}(\Delta)}{h_{y_0}(\Delta)} \,  \nu^{(n)}(\D)
\ee
 converges to some limiting Radon measure $\nu^{(n)}{}' \to \nu'(\Delta)$, in the weak$^*$ topology.
 But since $\nu^{(n)}{} \to \nu(\Delta)$, then we also have $\nu^{(n)}{}' \to h_{y_0'}(\Delta)/h_{y_0}(\Delta) \,  \nu(\Delta)$ at least in the weak$^*$ topology induced by the \textbf{compactly supported} continuous functions on $[\Delta_*,\infty)$. We only get this weaker statement because we have multiplied by the possibly unbounded function $h_{y_0'}(\Delta)/h_{y_0}(\Delta)$. By uniqueness of limits, and the fact that integrals on compact regions uniquely determine Radon measures, this implies $\nu'(\Delta)= h_{y_0'}(\Delta) / h_{y_0}(\Delta)\,  \nu(\Delta)$. Hence the measures $a(\Delta)$ and $a'(\Delta)$ defined using $y_0$ and $y_0'$ coincide,
\be
a'(\Delta) = a(\Delta) \ .
\ee

We have therefore shown that $F(y) = \int \ud a(\D)\, F_\D(y)$ for all $y\in \mathbb{D}$. The integral converges absolutely because by \rf{longeq} it can be written as $\int \ud\nu(\Delta)\, \frac{F_\D(y)}{h_{y_0}(\D)}$ for $|y|<y_0$. Since $\nu(\Delta)$ is a positive Radon measure on $[\Delta_*,\infty)$ and $\frac{F_\D(y)}{h_{y_0}(\D)}$ a continuous function vanishing at infinity, this integral converges absolutely. The same argument applies with $G_\D$ in place of $F_\D$ to conclude that $\int \ud a(\D)\, G_\D(y)$ converges absolutely at each $y\in \mathbb{D}$.
\end{proof}
Note that this argument could not be consistently restricted to discrete measures: they do not have the same sequential compactness property because discrete measures of bounded mass can accumulate into continuous ones.

\paragraph{Fenchel-Rockafellar optimisation.}
It is difficult to write down the objective function $a_{\Delta_*}$, since this is not a well-defined, let alone continuous, functional of $F=\int \ud a(\Delta)\, F_\Delta$. The issue of well-definedness is because decompositions into crossing vectors are not unique: one can change them by adding any solution of crossing. Well-definedness can be addressed by asking to maximise $a_{\Delta_*}$ such that $F\in A^2$ admits \textbf{some} decomposition  
\be
F = S^{(-)} + a_{\Delta_*} F_{\Delta_*} + \int_{(\Delta_*,\infty)}\ud a(\Delta)\, F_\D \ , \la{decF}
\ee
where $S^{(-)}$ is the crossing-antisymmetric part of the source below the gap,
\be
S^{(-)}(z) = S(z)-S(1-z) \ .
\ee
Recall that, in the simplest case, the source is just the identity block $S = G_0$, so $S^{(-)} = F_0$.

However, this is not a continuous objective function, as the value $a_{\Delta_*}\equiv a_{\{\Delta_*\}}$ of the atom at $\Delta_*$ can jump along convergent sequences. For example, it jumps from $0$ to $1$ at the limit point of $F_{\Delta_*+n^{-1}} \to F_{\Delta_*}$.

When combined with positivity, however, this can be formulated as a continuous objective function using the following trick. We shall append an extra co-ordinate $t \in \mathbb{R}$ to keep track of $a_{\Delta_*}$, and we will work in the Hilbert space $\mathbb{R}\oplus A^2$. Now, instead of imposing something like $t=a_{\Delta_*}$, we will only impose that $t\leq a_{\Delta_*}$. This is actually a closed constraint when imposed simultaneously with positivity, because limits of blocks above the gap can only accumulate to \textit{increase} the value $a_{\Delta_*}$ at the gap, preserving the inequality. This might seem like changing the optimisation problem, but it is in fact an equivalent problem, since maximising $t$ will simply select the largest possible value $t=a_{\Delta_*}$. This turns out to be a rather standard manipulation of objective functions, called the epigraph/hypograph trick (see e.g.\ \cite{vandenberghe2004convex}).

We therefore define the \textit{admissible set} to be
\be
\mathcal A := \left\{\ (t,F)\in\mathbb{R}\oplus A^2 \ \  \middle|  \ \ F = S^{(-)} + a_{*} F_{\Delta_*} + F_+ \  ,  \ \ 0\leq t\leq a_{*} \ ,  \ \  F_+ \in \overline{C_+}  \ \right\} \ .
\ee
We are denoting by $a_*$ the explicit coefficient of $F_{\Delta_*}$, but note that $F_+$ may be decomposed  in a way that contains the atom at ${\Delta_*}$ with any positive coefficient. Therefore, in decompositions of $F$ of the form \rf{decF}, the full coefficient $a_{\Delta_*}$ of this atom can be any number equal to $a_*$ or greater. We have thus arranged that $t\leq a_*\leq a_{\Delta_*}$.
\needspace{2\baselineskip}
\begin{lemma}
  $\mathcal A$ is closed and convex.
\end{lemma}
\begin{proof}
    It is clear that membership of $\mathcal A$ is preserved by convex combinations. 

For closedness, suppose elements $(t^{(n)},F^{(n)})\in\mathcal A$ converge to some $(t,F)$ in $\mathbb{R}\oplus A^2$. Writing $F^{(n)} = S^{(-)}+a^{(n)}_{\Delta_*} F_{\Delta_*} + F_+^{(n)}$ as above and evaluating the derivative at $z=\ha$, we find
\be
0 \leq a^{(n)}_{\Delta_*} F_{\Delta_*}'(z=\ha) \leq F^{(n)}{}'(z=\ha)-S^{(-)}{}'(z= \ha)\ ,
\ee
because $F_{\Delta_*}'(z=\ha)$ and $F_+^{(n)}{}'(z=\ha)$ are non-negative by Lemma~\ref{lemma:derivativePos}. Hence $a^{(n)}_{\Delta_*}$ are a bounded positive sequence, so they have a  subsequence converging to a limit $a_{*}$, which must therefore satisfy $0\leq t\leq a_{*}$. Since $F^{(n)}\to F$ then $F_+^{(n)}\to F_+ = F-S^{(-)}-a_{*}F_{\Delta_*}$. We must have $F_+ \in \overline{C_+}$ since it is closed. Therefore the limit $(t,F)\in\mathcal A$.
\end{proof}

\begin{problem}[OPE-max in Hilbert space]\label{prob:HOPE}
    Define $f,g:\mathbb{R} \oplus A^2 \to \mathbb{R}\cup \{+\infty\}$ by 
    \begin{align}
        f(t,F) = -t + 1_{\mathcal A}(t,F) \ , \qquad \qquad g(t,F) = 1_{\mathbb{R}\oplus\{0\}}(t,F) \ . 
    \end{align}
    \textbf{Primal:} \qquad Minimise \quad \quad  $f+g$ .
\end{problem}
\noindent The functions $f$ and $g$ are convex: $t$ is linear (so convex) and the indicator functions are on convex sets. They are also lsc: $t$ is clearly continuous and the constraint sets $\mathcal A$ and $\mathbb{R}\oplus\{0\}$ are closed. They are proper functions because these constraint sets are clearly non-empty.

This formulation of the problem may seem abstract, but ultimately it is just maximising the OPE coefficient $a_{\Delta_*}$ subject to crossing and unitarity. The following result shows that it is equivalent to the physical OPE-max Problems~\ref{prob:opemax} and \ref{prob:Gopemax}.
\begin{proposition}[equivalence of problems]\label{prop:problemEquivalence}
    The Hilbert space OPE-max Problem~\ref{prob:HOPE} is {\rm equivalent} to the physical OPE-max Problem~\ref{prob:Fopemax} for Radon measures $a(\Delta)$. This means that for any feasible point of one problem, the other problem has a feasible point with the same or higher objective function. In particular, if one problem has optimal attainment, then so does the other.
\end{proposition}
\begin{proof}
This follows easily from the preceding discussion. On one hand, if a Radon measure $a(\Delta)$ is feasible for the physical problem (a unitary solution of crossing), then by Lemma~\ref{lemma:OPEA2}, its OPE converges in $A^2$, and hence so does $\int \ud a(\D) \, F_\D=0$. Therefore $(a_{\{\D_*\}},0) \in \mathbb{R}\oplus A^2$ is feasible for the Hilbert space problem with the same value of the objective function.

Conversely, if $(t,F)$ is feasible for the Hilbert space problem, then by Lemma~\ref{lemma:clcone}, $F(y)=S^{(-)}(y)+a_* F_{\Delta_*}(y)+ \int_{[\Delta_*,\infty)} \ud b(\D) \, F_\D(y)=0$ for some positive Radon measure $b(\D)$ and $0\leq t\le a_*$, with the integral converging absolutely  at each $y\in\mathbb{D}$. As well as the sum of crossing vectors, the OPE sum of blocks $G(y)=S(y)+a_*G_{\Delta_*}(y)+\int db(\Delta)\,G_\Delta(y)$ converges absolutely at each $y$, because blocks are dominated by functional $h_{y_0}(\Delta)$ with $y_0>|y|$, by Lemma~\ref{lemma:posFunc}. This is a positive solution of crossing, i.e.\ a feasible solution for the physical problem with objective function $a_{\{\Delta_*\}} = a_*+b_{\{\Delta_*\}}\geq t$ i.e.\ at least as big as that of $(t,F)$.

Finally, it follows that both or neither problem must have optimal attainment. Suppose $x$ is an optimal solution for one of the problems---call it (a). Then the other problem, (b), has a solution $y$ at least as good as $x$. It cannot be strictly better than $x$, since otherwise the problem (a) would have to have a solution strictly better than $x$. For the same reason, there can be no solution of (b) strictly better than $y$. Hence $y$ is an optimal solution of (b).
\end{proof}

\paragraph{The dual problem.} As pointed out in Section~\ref{convOpt}, the split of the objective function into $f$ and $g$ is arbitrary, but it can drastically modify the dual problem. In our case, for example, one could shuffle any combinations of the three terms $-t$, $1_{\mathcal A}$ and $1_{\mathbb{R}\oplus\{0\}}$ into $f$ and $g$. We find that the present choice, and the one with $1_{\mathcal A}$ and $1_{\mathbb{R}\oplus\{0\}}$ swapped, lead to the same dual problem, which is a formalised version of the familiar dual problem \rf{informalDual}. In contrast, the other combinations lead to trivial duals with only one feasible point (so the interiority condition \rf{GSlater2} is certainly  not satisfied).

Let us write down the convex dual of the primal problem \ref{prob:HOPE}. We are working in the real Hilbert space $X=\mathbb{R}\oplus A^2$ with the inner product\footnote{Since it is always clear from the arguments, we use the same notation $\langle-|- \rangle$ for the inner products of $A^2$ and $\mathbb{R}\oplus A^2$.}
\begin{equation}
    \langle (s,H)\, |\, (t,F)\rangle = st+\langle H\ket{F}\ .
\end{equation}
The convex dual \eqref{FenchelDual} of the function $g(t,F)=1_{\mathbb{R}\oplus \{ 0 \}}(t,F)$ is given by
\begin{equation}\begin{split}
    g^*(s,H) = \sup_{t\in\mathbb{R}}\big\langle(s,H),(t,0)\big\rangle = \sup_{t\in \mathbb{R}} \left(st\right)&=1_{\{0\}\oplus A^2}(s,H) \\
    &=
    \begin{cases}
        0  &, \qquad s=0 \ ,\\
        +\infty  &,\qquad s\neq 0\ .
    \end{cases}\label{gConjugate}
\end{split}\end{equation}
For $f(t,F) = -t+1_{\mathcal{A}}(t,F)$ we have
\begin{equation}
    f^*(s,H) = \sup_{(t,F) \in \mathcal A} \big((s+1)t +\langle H\ket{F}\big)\ .
\end{equation}
Elements of $\mathcal A$ can be parametrised (non-uniquely) as $F= S^{(-)}+a_*F_{\D_*}+F_+$, where $0\leq t\leq a_*$ and $F_+\in \overline{C_+}$ range independently. Hence, writing $\omega_H[F] = \langle H\ket F$, we may decompose
\begin{equation}
  f^*(s,H) = \omega_H[S^{(-)}] + \sup_{\substack{a_*\geq0\\ 0\leq t\leq a_*}}\big((s+1)\, t+a_*\, \omega_H[F_{\D_*}]\big) + \sup_{F_+\in \overline{C_+}}\omega_H[F_+]\ . 
\end{equation}
Starting with the last term, if $\omega_{H}[F_{\Delta}]>0$ for any $\Delta\geq\Delta_*,$ then  $\lambda F_{\D}\in \overline{C_+}$ with $\lambda\to\infty$ has unbounded value $\omega_H[\lambda F_\D] \to +\infty $ so the supremum is $+\infty.$ On the other hand, suppose $\omega_H[F_{\Delta}]\leq0$ for all $\Delta\geq\Delta_*$. Then it follows by continuity of $\omega_H$ that $\omega_H[F_+]\leq0$ for all $F_+\in\overline{C_+}$, so the supremum is $0$ and is  attained by $F_+ = 0$. The last term is therefore $1_{\omega_H(\Delta)\leq 0,\ \D\geq \D_*}(s,H)$.

To evaluate the middle term, first maximising over $t$ gives $a_{*}(\max(s+1,0)+\omega_H[F_{\Delta_*}])$. Then taking the supremum over $a_{*}$ gives $0$ if $\max(s+1,0)+\omega_H[F_{\Delta_*}]\leq0$, and $+\infty$ otherwise. Since we already have the constraint $\omega_H[F_{\Delta_*}]\leq0$ from the last term, this is equivalent to $s+1+\omega_H[F_{\D_*}]\leq0$. Collecting, we have
\begin{equation}
    f^*(s,H) = \omega_H[S^{(-)}]+1_{\mathcal{P}}(s,H)\label{fConjugate}
\end{equation}
where 
\begin{equation}
    \mathcal{P} :=\left\{(s,H)\ \middle|\ \omega_H(\Delta)\leq0 \text{ for } \Delta\geq \Delta_* \text{ and } \omega_H(\Delta_*)\leq -1-s\right\}. \label{DualCone}
\end{equation}
\needspace{2\baselineskip}
The dual problem is therefore:
\begin{problem}[convex dual of OPE-max]\label{prob:DualHOPE}
    Define $f^*,g^*:\mathbb{R} \oplus A^2 \to \mathbb{R}\cup \{+\infty\}$ by 
    \begin{align}
        f^*(s,H) = \omega_H[S^{(-)}] + 1_{\mathcal P}(s,H) \ , \qquad \qquad g^*(s,H) = 1_{\{0\}\oplus A^2}(s,H) \ . \label{fgdual}
    \end{align}
 \textbf{Dual:}  \qquad  Maximise   \quad \quad  $-f^*(s,H)-g^*(-s,-H)$\ .
\end{problem}

This is equivalent to the informal dual problem \rf{informalDual}, over the class of Bergman functionals. To see this, first note that $g^*$ imposes the constraint $s=0$, under which $f^*$ simplifies to
\begin{equation}
    f^*(0,H) = \begin{cases}
        \omega_H[S^{(-)}] \quad &\text{ if } \omega_H(\Delta)\leq0 \text{ for } \Delta\geq \Delta_* \text{ and } \omega_H(\Delta_*)\leq -1 \ ,\\
        +\infty &\text{ otherwise}\ ,
    \end{cases}
\end{equation}
so the dual problem is then equivalently written as
\begin{equation}
    D
     = \sup_{(s,H)\in X}\big(-f^*(s,H)-g^*(-s,-H)\big) 
    = \sup_{\substack{\omega_{H}(\Delta_*)\leq-1\\ \omega_{H}(\Delta)\leq0}}\big(-\omega_{H}[S^{(-)}]\big) \ .
\end{equation}
This is the same problem as \rf{informalDual} upon identifying $\omega=-\omega_H$.

\subsection{Generalised Slater condition (proof of Theorem~\ref{thm:main})}
We shall now complete the proof of Theorem \ref{thm:main} by applying part (ii) of the Fenchel-Rockafellar Theorem \ref{thm:FR}. As explained above, the functions $f,g$ are proper, convex and lsc as needed for the theorem.

Let us study the regularity condition \rf{GSlater2}. We need to show that $0$ is in the interior of the domain sum $\operatorname{dom}f^* +\operatorname{dom}g^*$ of the dual problem. More strongly than  this, we claim that $\operatorname{dom}f^* +\operatorname{dom}g^* = \mathbb{R} \oplus A^2$ is the whole space, so $0$ is certainly in its interior.

To see that, we read off the domains from \rf{fgdual},
\begin{equation}
      \operatorname{dom}f^* = \mathcal P \ , \qquad \operatorname{dom}g^* = \{0\}\oplus A^2 \label{eq:doms}
\end{equation}
to find that the second entry covers all of $A^2$:
\be
\operatorname{dom}f^* +\operatorname{dom}g^* = \mathcal P_{\mathbb{R}}\oplus A^2 \ ,
\ee
where we define the projection onto the first element,
\be
\mathcal P_{\mathbb{R}} = \left\{ s \ \middle| \ (s,H)\in \mathcal P \text{ for some }H\in A^2\right\} \ .
\ee
We will now show that $\mathcal{P}_\mathbb{R}=\mathbb{R}$, i.e.\ all numbers appear in the first entry of $\mathcal P$. This will follow from the existence of a positive functional $\omega$.  We already showed in Lemma~\ref{lemma:posFunc} that strictly positive functionals exist: we can take $\omega = \omega_{h_{y_0}}$ defined there for any $y_0\in (0,1)$. Equivalently, $(0,-\lambda \, h_{y_0})$ is an element of $\mathcal P$ for some $\lambda>0$.\footnote{Note that it would actually be good enough for this argument for $\omega$ to be strictly positive at $\Delta_*$ and weakly positive above $\Delta_*$.} Now by sufficiently rescaling this functional by $\lambda \to |1+s|\lambda$, we can hit all numbers $s\in \mathbb{R}$ in the first entry of elements of $\mathcal P$. Indeed, since we currently have $-\lambda \, \omega(\Delta_*)\leq -1$, then certainly $- |1+s|\lambda \, \omega(\Delta_*)\leq -|1+s| \leq -1-s$. Hence $(s,-|1+s|\lambda\, h_{y_0})\in \mathcal P$ for all $s\in \mathbb{R}$. This proves that the regularity condition \rf{GSlater2} is satisfied.

The Fenchel-Rockafellar theorem therefore applies,  so we have zero duality gap $P=D$, and the primal optimum $P$ is attained if it is finite. Because of our assumption $\Delta_*\geq \sqrt{2}\Delta_\phi$, we know that $P>-\infty$, i.e. the primal problem is bounded. We have therefore proved the statement of Theorem~\ref{thm:main}: if the primal problem is feasible ($P<+\infty$), then its optimal value is attained.

\section{Discussion}\label{sec:disc}
In 1d CFT OPE maximisation problems of a single correlator, we have given a rigorous proof that an extremal solution exists if any feasible solution exists. These are unitary, exact solutions of crossing. Moreover, we expect that generic cases will be non-trivial and interacting. Our method for proving this was to take the problem seriously as a convex optimisation problem in an infinite dimensional space. We emphasise, however, that we have not just proven formal existence in some abstract space, but the existence of actual extremal correlators of the form 
\be
G(z) = \int \ud a(\Delta)\,  G_\Delta(z) \ .
\ee
In particular, the OPE decomposition is baked into our choice of feasible set.

The low-lying source $S(z)$ was an arbitrary positive combination of blocks below the gap $\Delta_*$, with the gap assumption that the correlators take the form
\be
G(z) = S(z) + \int_{[\D_*,\infty)} \ud a(\Delta) \, G_\D(z) \ . \label{GSgap}
\ee
While we assumed the natural form $S(z) = \int_{[0,\Delta_*)}\ud a^{\rm fixed}(\Delta) \, G_\D(z)$ above, in fact we believe a similar argument goes through with $S(z)$ taken to be any element of $A^2$ with the property $|S^{(-)}(z)|=O(z^{-2\Delta_\phi})$ as $z\to0$ along the real axis $z>0$, and $a(\Delta)$ remaining a positive Radon measure on $[\D_*,\infty)$. For example, $S(z)$ does not need to consist only of blocks in the interval $[0,\D_*)$ below the gap. Moreover, it may contain blocks with \textbf{negative} weight: this could be an interesting way to tackle non-unitary theories by injecting non-positivity.

\paragraph{Extremal functionals.} We also showed that the primal and dual problems have zero duality gap. This means there are dual functionals giving bounds arbitrarily close to being saturated by the extremal correlators. We have not yet been able to prove dual attainment, i.e.\ the existence of extremal functionals. Our numerical explorations strongly suggest, at least in simple examples, that the dual does attain its optimal value in $A^2$ if the parameters $\alpha$ and $\beta$ are close enough to their maximum values in \eqref{eq:gamma-range}. This is why we have chosen to keep these parameters free, rather than setting them e.g. to the simplest choice $\alpha=\beta=0$, which would actually prevent dual attainment according to our numerics.

Analytically proving dual attainment is a major goal for future work. A natural approach would be to try to apply part (i) of the Fenchel-Rockafellar Theorem~\ref{thm:FR}---i.e.\ to argue as above with the primal and dual swapped. However, we believe that this particular interiority condition (or the simplest generalisations of it, see \cite{boct2011comparison}) is not satisfied in this case. Since this is only a sufficient condition for dual attainment, its failure does not say anything about whether the dual optimum is actually attained.

\paragraph{Discreteness and sparseness.} In this article, we proved that OPE-maximising extremal solutions exist, at least with Radon measures as their OPE densities. This leaves open the possibility of continuous spectra. However, based on general empirical observations \cite{Paulos:2019fkw,part2,Ghosh:2025sic}, we expect the extremal solutions to resemble Generalized Free Field theories, and in particular to have discrete, sparse spectra. In fact, dual attainment offers a route to proving this by the following mechanism, called \textit{complementary slackness}. If one has primal and dual attainment with zero duality gap, then the inequality in \rf{OPEbound} must be saturated by them. That means the sum rule receives zero contribution from above the gap, $\int_{(\Delta_*,\infty)}\ud a(\Delta) \, \omega(\Delta) = 0$. But the integrand is positive, so $a(\Delta)$ must be supported at the zeros of $\omega(\Delta)$: this is the familiar statement that the extremal spectrum is contained in the set of zeros of the extremal functional.
Under general assumptions $\omega(\Delta)$ can be argued to be analytic (see \cite{part2}), so its zeros must be discrete. One concludes that the extremal spectrum, i.e.\ the support of $a(\Delta)$, is discrete. Moreover, with further analysis of the properties of $\omega(\Delta)$, it may be possible to derive the asymptotic density of its zeros, and hence the behaviour of the extremal spectrum at large $\D$.

\paragraph{Uniqueness.} It would be interesting to study whether the extremal solutions of this paper are unique. This seems to be an expectation of the bootstrap literature \cite{ElShowk:2012}, although it has not been rigorously established in our setting.

\paragraph{$\Lambda\to\infty$ convergence.} 
In practice in the numerical conformal bootstrap, one truncates these problems to finite-dimensional ones. On the functional side, the standard truncation is to a finite number $\Lambda$ of derivative functionals, which form a basis for a finite-dimensional subspace of functionals. It would be interesting to study the convergence of these truncations as $\Lambda\to\infty$. In particular, we expect that the optimal values of the objective functions converge in this limit.

\paragraph{Other conformal bootstrap problems.} It would be interesting to generalise this study to the $d>1$ conformal bootstrap. This entails handling two independent cross-ratios $(z,\bar z)$ and conformal blocks that are less explicit in odd $d$. We do not anticipate any conceptual obstructions in that exercise.

It would also be  useful to consider another central class of bootstrap problems: \textbf{gap maximisation}. Applying these methods there is not completely straightforward, as the objective function in that case is not convex. However, there are several natural approaches for dealing with this. First, we believe that convexity is not an essential element of this approach, and that it can be reformulated in terms of more general compactness arguments. Second, one can still try to construct a convex optimisation problem that acts as a proxy for gap maximisation. Minimisation of the 4-point correlator has been proposed as such a proxy \cite{Paulos:2020zxx,Paulos:2021jxx}. 

For practical applications, it would be important to apply these methods to problems with multiple correlation functions, i.e.\ genuine semi-definite programs, as opposed to linear programs. As mentioned above, such problems already encounter non-trivial questions of Slater's condition and optimal attainment in finite dimensions.

\paragraph{The Bergman space.} We found the weighted Bergman space a useful and natural setting for this work. It has several advantageous properties: (i) it is a Hilbert space, (ii) it models both correlation functions and functionals with the same space, and (iii) its class of functionals is very large, containing previously studied families. We anticipate that this space may be useful for addressing the further questions discussed above. It could also be useful for numerics to consider expansions of functionals into bases for the Bergman space. These may offer an alternative to the analytic functionals used in \cite{Ghosh:2023onl,Ghosh:2026dse} that could be cheaper to evaluate.

\section*{Acknowledgments}
We thank Rajeev Erramilli, Petr Kravchuk, Dalimil Maz\'a\v{c}, Miguel Paulos, Marten Reehorst and Roman Shabanov for useful discussions and comments on the draft. NL particularly thanks Slava Rychkov for a discussion that inspired his interest in this topic. NL is supported by the European Research Council (ERC) under the European Union’s Horizon 2020 research and innovation programme (grant agreement No 101115511).

\vspace{0.2cm}
\noindent \textbf{AI use statement:} We acknowledge the use of Claude (Anthropic) and ChatGPT (OpenAI) to check the validity of arguments and search the literature. The authors take the full responsibility for the contents of the paper.

\bigskip
\appendix
\addtocontents{toc}{\protect\setcounter{tocdepth}{1}}
\section{Derivations of properties of blocks}\label{App:Blocks}
In this appendix, we provide proofs of Lemmas \ref{lemma:RadDom}--\ref{lemma:derivativePos}.
\subsection{Proof of Lemma \ref{lemma:RadDom}}
We begin by showing that $|\rho(y)|\leq \rho(|y|).$ To do this, we set $x=\sqrt{\rho}$, choosing the principal branch with $x$ valued in the
right half unit disc.
We can solve the relations
\begin{equation}
    z(y) = \frac{(1+y)^2}{2(1+y^2)} \ , \qquad  \quad z(\rho) = \frac{4\rho}{(1+\rho)^2}
\end{equation}
to obtain the following relation between $y$ and $x=\sqrt{\rho}$,
\begin{equation}
    y(x) = \frac{x-(\sqrt2-1)}{x+(\sqrt{2}-1)}\frac{1+(\sqrt2-1)x}{1- (\sqrt{2}-1)x}\ .\label{A2}
\end{equation}
It is useful to introduce $A= |x|^2+(\sqrt2-1)^2,\ B=1+(\sqrt2-1)^2|x|^2$ and $t=2(\sqrt2-1)\, \Re(x)$, in terms of which we can write
\begin{equation}
    |y(x)|^2=\frac{(A-t)(B+t)}{(A+t)(B-t)}\ .
\end{equation}
This depends on the complex variable $x$ through the two real components $(|x|,t)$. Holding $|x|$ fixed and taking the derivative with respect to $t$, we find
\begin{equation}
   \frac{\partial}{\partial t} |y(x)|^2 = -\frac{2(B-A)(AB+t^2)}{(AB-t^2+(B-A)t)^2}<0
\end{equation}
since $B-A = (1-(\sqrt{2}-1)^2)(1-|x|^2)>0$. Therefore, on each circle of constant $|x|$, $|y(x)|$ attains its minimum value at $x=|x|$,
\begin{equation}
    |y(|x|)| \leq |y(x)| \ .\label{A1:boundOnY}
\end{equation}
Since $y(|x|)$ is real then this implies
\be
y(|x|) \leq |y(|x|)|\leq |y(x)|= y(x(|y(x)|)) \ .
\ee
Both $|x|$ and $x(|y(x)|)$ are in $[0,1)$. We can invert this inequality by using the fact that $y(r)$ is an increasing function on $r\in[0,1)$,
\begin{equation}
    y'(r) = \frac{4(1+r^2)}{(1+2r-r^2)^2}>0 \ .
\end{equation}
Thus we have
\be
|x(y)| \leq x(|y|) \ .
\ee
Re-writing this in terms of $\rho=x^2$ gives
\be
|\rho(y)|\leq \rho(|y|)
\ee
as claimed.

Consider now 
\begin{align}
    \widehat{G}_{\Delta}(\rho) := z(\rho)^{2\Delta_{\phi}}G_{\Delta}(z(\rho))&= {(4\rho)^{\Delta}} _2F_1(\tfrac{1}{2},\Delta;\Delta+\tfrac{1}{2};\rho^2)\\
    &=(4\rho)^{\Delta}\sum_{n=0}^{\infty}\frac{(\tfrac12)_{n}(\Delta)_n}{n!(\Delta+\tfrac12)_n}\rho^{2n}\label{A1:SeriesExpansionBlocks}
\end{align}
where in the first line we used the quadratic transformation of the hypergeometric function. For $\Delta\geq0$ the coefficients of the series in the second line are all non-negative, and hence, using $|\rho(y)|\leq \rho(|y|)$ we have
\begin{equation}
    |\widehat G_\Delta(\rho(y))|\leq \widehat G_{\Delta}(|\rho(y)|)\leq \widehat G_{\Delta}(\rho(|y|)).
\end{equation}
Restoring the external dimension prefactors, we have
\begin{equation}
    |G_{\Delta}(y)|\leq\bigg|\frac{z(|y|)}{z(y)}\bigg|^{2\Delta_{\phi}}G_{\Delta}(|y|)
\end{equation}
\subsection{Proof of Lemma \ref{lemma:ratBlocks}}
These are very standard properties. Let us start from the expansion \eqref{A1:SeriesExpansionBlocks} of conformal blocks in the $\rho$ variable,
\begin{equation}
    G_{\Delta}(z)=  z^{-2\Delta_{\phi}}\,  (4\rho(z))^{\Delta}\sum_{n=0}^{\infty}\frac{(1/2)_n}{n!}\frac{(\Delta)_n}{(\Delta+1/2)_n}\rho(z)^{2n}\ . \label{rhoExpBlock}
\end{equation}
At $0<z<1$, every term in the series is positive, so $G_{\Delta}(\rho)>0$.

Additionally, defining $S_\Delta(r)= \sum_{n=0}^{\infty}\frac{(1/2)_n}{n!}\frac{(\Delta)_n}{(\Delta+1/2)_n} r^{2n}$, then for $0<z_1<z_2<1$ we have 
\begin{equation}
    \frac{G_{\Delta}(z_1)}{G_{\Delta}(z_2)} =  \bigg(\frac{z_2}{z_1}\bigg)^{2\Delta_{\phi}}\left(\frac{\rho(z_1)}{\rho(z_2)}\right)^{\Delta}\frac{S_\D(\rho(z_1))}{S_\D(\rho(z_2))}\ .\label{A2:RatioOfBlocks}
\end{equation}
Now note that $S_\Delta(r)$ is increasing in $r\in[0,1)$ because it is defined by a Taylor series with positive coefficients. Hence we have
\begin{equation}
    \frac{G_{\Delta}(z_1)}{G_{\Delta}(z_2)} \leq  \bigg(\frac{z_2}{z_1}\bigg)^{2\Delta_{\phi}}\left(\frac{\rho(z_1)}{\rho(z_2)}\right)^{\Delta}\ ,
\end{equation}
which decays exponentially as $\Delta\to\infty$.

\subsection{Proof of Lemma \ref{lemma:crossPos}}
We start by considering the block given by
\begin{equation}
    G_{\Delta}(z) ={z^{\Delta-2\Delta_{\phi}}}_2F_1(\Delta,\Delta;2\Delta;z) = {z^{\Delta-2\Delta_{\phi}}}\sum_{n=0}^{\infty}\frac{(\Delta)_n^2}{n!(2\Delta)_n}z^n
\end{equation}
Since each coefficient is positive and $z^{\Delta-2\Delta_{\phi}}$ is non-decreasing for $\Delta\geq2\D_{\phi}$, then $G_{\Delta}(z)$ is a strictly increasing function for $0<z<1$. Hence, for $1/2<z<1$ we have
\begin{equation}
    0<G_{\Delta}(1-z)<G_{\Delta}(z)
\end{equation}
and therefore
\begin{equation}
    0<F_{\Delta}(z)=G_{\Delta}(z)-G_{\Delta}(1-z)<G_{\Delta}(z)\ .
\end{equation}
To strengthen the lower bound, we define
\begin{equation}
    \k_{z} :=\sup_{\Delta\geq2\Delta_{\phi}}\frac{G_{\Delta}(1-z)}{G_{\Delta}(z)}.
\end{equation}
We know that the ratio inside the supremum is positive, continuous in $\Delta$ and strictly less than $1$. Moreover, by Lemma~\ref{lemma:ratBlocks} we know that for $1/2<z<1$,
\begin{equation}
   \frac{G_{\Delta}(1-z)}{G_{\Delta}(z)}\rightarrow 0 \quad \text{ as }\Delta\rightarrow\infty\ ,
\end{equation}
so the supremum must be attained at a finite value of $\Delta$. But since the ratio is strictly less than $1$, this implies $\k_z\in(0,1)$. Thus
\begin{equation}
    G_{\Delta}(1-z)\leq \k_z\, G_{\Delta}(z)\ ,
\end{equation}
and hence
\begin{equation}
    (1-\k_z)G_{\Delta}(z)\leq F_{\Delta}(z)\leq G_{\Delta}(z)\ .
\end{equation}

\subsection{Proof of Lemma \ref{lemma:derivativePos}}
At the crossing-symmetric point $z=\ha$, we have $F_\Delta'(\ha)=2G_\Delta'(\ha)$. Let us use the expression \eqref{rhoExpBlock} for the block. Taking its logarithmic derivative, we find
\be
\frac{G_\D'(z)}{G_\D(z)} = \partial_z \log\left(z^{-2\D_\phi}\right) + \partial_z \log\left(\rho(z)^\D\right) + \partial_z \log\left(\sum_{n=0}^\infty \frac{(\ha)_n(\D)_n}{n!(\Delta+\ha)_n}\, \rho(z)^{2n}\right) 
\ee
The argument of $\log$ in the last term is a strictly positive function with strictly positive derivative on $z\in(0,1)$, so that term is positive:
\be
\frac{G_\D'(z)}{G_\D(z)} > \partial_z \log\left(z^{-2\D_\phi}\right) + \partial_z \log\left(\rho(z)^\D\right) = \frac{-2\D_\phi}{z} + \frac{\D}{z\sqrt{1-z}} \ .
\ee
At $z=\ha$, this gives 
\be
G_\D'(\ha) > 2\sqrt{2} \left(\D-\sqrt{2}\Delta_\phi\right)G_\D(\ha) 
\ee
and hence $F_\Delta'(\ha)=2G_\Delta'(\ha)>0$, as desired, if $\D\geq \sqrt{2}\D_\phi$.

\bibliography{bib}
\bibliographystyle{JHEP}

	\end{document}